\documentclass[journal]{IEEEtran}

\usepackage{graphicx,psfrag,epsfig,epsf,latexsym,hhline,amsmath,amssymb,multirow}
\usepackage{pst-plot}
\usepackage{pstricks-add}
\usepackage{pifont}
\usepackage{graphicx}
\usepackage{amsthm,bm}
\usepackage[linesnumbered,ruled,vlined]{algorithm2e}
\usepackage[noadjust]{cite}
\usepackage{array}
\usepackage{blindtext}
\usepackage{etoolbox}
\usepackage{comment}
\usepackage{epstopdf}
\usepackage{dsfont}
\usepackage[flushleft]{threeparttable}
\usepackage{hyperref}
\hypersetup{
    unicode=false,          
    pdftoolbar=true,        
    pdfmenubar=true,        
    pdffitwindow=false,     
    pdfstartview={FitH},    
    pdftitle={My title},    
    pdfauthor={Author},     
    pdfsubject={Subject},   
    pdfcreator={Creator},   
    pdfproducer={Producer}, 
    pdfkeywords={keyword1, key2, key3}, 
    pdfnewwindow=true,      
    colorlinks=true,       
    linkcolor=blue,          
    citecolor=blue,
    filecolor=cyan,         
    urlcolor=magenta        
}

\SetCommentSty{mycommfont}

\SetKwInput{KwInput}{Input}                
\SetKwInput{KwOutput}{Output}              

\newcolumntype{M}[1]{>{\centering\arraybackslash}m{#1}}
\newcolumntype{P}[1]{>{\centering\arraybackslash}p{#1}}

\newcommand{\msf}[1]{\mathsf{#1}}

\newcommand{\SNR}{\text{SNR}}

\newcommand{\E}{\mathbb{E}}

\newcommand{\iid}{i.\@i.\@d.\ }

\theoremstyle{plain}
\newtheorem{lemma}{Lemma}

\newtheorem{theorem}{Theorem}
\newtheorem{remark}{Remark}

\newcommand\xqed[1]{%
  \leavevmode\unskip\penalty9999 \hbox{}\nobreak\hfill
  \quad\hbox{#1}}
\newcommand\demo{\xqed{$\blacksquare$}}

\IEEEoverridecommandlockouts
\begin{document}
\title{Uplink Multiple Access with Heterogeneous Blocklength and Reliability Constraints: Discrete Signaling with Treating Interference as Noise}
\author{Min Qiu,~\IEEEmembership{Member,~IEEE}, Yu-Chih Huang,~\IEEEmembership{Senior Member,~IEEE}, and Jinhong Yuan,~\IEEEmembership{Fellow,~IEEE}

\thanks{This work has been presented in part at the 2023 IEEE Global Communications Conference (GLOBECOM), Kuala Lumpur, Malaysia \cite{Qiu23}.

The work of Min Qiu and Jinhong Yuan was supported in part by the Australian Research Council (ARC) Discovery Project under Grant DP220103596, and in part by the ARC Linkage Project under Grant LP200301482. The work of Yu-Chih Huang was supported by the National Science and Technology Council, Taiwan, under Grant NSTC 113-2223-E-A49-005-MY3. This work was also supported in part by the Higher Education Sprout Project of the National Yang Ming Chiao Tung University and Ministry of Education (MOE), Taiwan.

Min Qiu and Jinhong Yuan are with the School of Electrical Engineering and Telecommunications, University of New South Wales, Sydney, NSW, 2052 Australia (e-mail: min.qiu@unsw.edu.au; j.yuan@unsw.edu.au).
Yu-Chih Huang is with the Institute of Communications Engineering, National Yang Ming Chiao Tung University, Hsinchu 300, Taiwan (e-mail: jerryhuang@nycu.edu.tw).
}
}

\maketitle

\begin{abstract}
We consider the uplink multiple access of heterogeneous users, e.g., ultra-reliable low-latency communications (URLLC) and enhanced mobile broadband (eMBB) users. Each user has its own reliability requirement and blocklength constraint, and users transmitting longer blocks suffer from heterogeneous interference. On top of that, the decoding of URLLC messages cannot leverage successive interference cancellation (SIC) owing to the stringent latency requirements. This can significantly degrade the spectral efficiency of all URLLC users when the interference is strong. To overcome this issue, we propose a new multiple access scheme employing discrete signaling and treating interference as noise (TIN) decoding, i.e., without SIC. Specifically, to handle heterogeneous interference while maintaining the single-user encoding and decoding complexities, each user uses a single channel code and maps its coded bits onto sub-blocks of symbols, where the underlying constellations can be different. We demonstrate theoretically and numerically that the proposed scheme employing quadrature amplitude modulations and TIN decoding can perform very close to the benchmark scheme based on Gaussian signaling with perfect SIC decoding. Interestingly, we show that the proposed scheme does not need to use all the transmit power budget, but also can sometimes even outperform the benchmark scheme.
\end{abstract}

\begin{IEEEkeywords}
Multiple access, URLLC and eMBB coexist, finite blocklength coding, discrete modulations, treating interference as noise.
\end{IEEEkeywords}

\section{Introduction}\label{sec:intro}
Future generation wireless communication systems are expected to support a variety of heterogeneous services. These services include enhanced mobile broadband (eMBB) and ultra-reliable low-latency communications (URLLC) services as already classified in the fifth generation (5G) systems, and other diversified use cases such as Augmented Reality and eHealth \cite{9040264}. Existing networks based on the conventional ``one-size-fits-all'' design are difficult to meet the diverse requirements of heterogeneous services due to the lack of flexibility and scalability \cite{8004168,8685766,9097306}. To support customized service requirements, the idea of network slicing has been introduced to enable the coexistence of heterogeneous services in the same radio access network \cite{7470940,7499297}. The current proposal is to slice the network by allocating orthogonal resources, e.g., in the time/frequency domain, for each service type \cite{8004168}. Since different services are isolated from each other, their quality-of-service requirements are guaranteed. However, this approach could lead to low spectral and energy efficiency when the number of service types and devices becomes large.

Various research has been carried out to investigate efficient coexistence mechanisms for heterogeneous services with improved efficiency and user fairness. Notably, \cite{8476595,8647460,9364885,9562192,10190330} introduced non-orthogonal multiple access schemes for eMBB and URLLC coexistence in the uplink. These schemes stem from the classical studies on the Gaussian multiple access channel (GMAC) with homogeneous and infinite blocklength, where the entire capacity region can be achieved by superposition coding and successive interference cancellation (SIC) with time-sharing \cite{Cover:2006:EIT:1146355} or rate-splitting with partial SIC \cite{485709,9831440}.

One salient difference between the homogeneous and heterogeneous multiple access is that the decoding of a URLLC transmission block cannot leverage SIC owing to its stringent latency requirements \cite{8476595,9036072}. That is, the base station should always decode URLLC services first before other non-urgent services, e.g., eMBB services, regardless of channel conditions. Without SIC, URLLC users can suffer from a larger rate penalty than in the orthogonal-based slicing schemes \cite{8476595,10190330}. For the decoding of eMBB messages, the receiver can use SIC to cancel the URLLC interference or simply treat the received symbols that are interfered by URLLC packets as erased at the cost of some performance degradation \cite{8476595,9364885}. That said, since the blocklength is short in URLLC \cite{7529226,8541123}, performing SIC by decoding URLLC first can fail because the decoding error probability cannot be arbitrarily small with finite blocklength coding \cite{5452208,8345745,8933345,9036072}. In addition, even if we assume that SIC is feasible for decoding URLLC, e.g., using early decoding \cite{9838392,10304170}, performing SIC by decoding eMBB first then URLLC would require the decoding of eMBB services to achieve at least the same reliability as for URLLC services. However, this is difficult to achieve since the coding schemes for eMBB messages generally have an error floor much higher than the URLLC reliability requirements \cite{10502324}. Further, SIC can introduce extra decoding delay, complexity, and error propagation, which can become pronounced when the number of users is large. In light of the above, the non-orthogonal schemes relying on SIC may not fully address the challenges that arise from the coexistence of heterogeneous services; thereby, new proposals are called for.

Non-asymptotic bounds are required to characterize the achievable rate under finite blocklength coding. The landmark work \cite{5452208} introduced an easy-to-compute and tight approximation for the achievable finite blocklength coding rate under non-vanishing decoding error probability in Theorem 54 therein. However, the approximation is valid for the point-to-point additive white Gaussian noise (AWGN) channel. Whereas for multiuser channels, the derivation of finite blocklength achievable rates needs to consider the channel characteristics, desired signal and interference distributions, decoding methods, etc., rather than signal-to-interference-plus-noise ratio (SINR) only. \cite{7300429} extended the idea of \cite{5452208} to derive the achievable rate region of the two-user GMAC under finite blocklength constraints with second-order approximations. Specifically, the achievable rate was derived based on the assumptions that the input distributions are shell codes, i.e., codewords drawn from a power shell \cite[Sec. X]{6767457}, and under joint typicality decoding. Recently, \cite{9535162} generalized the results from \cite{7300429} to the $K$-user GMAC and refined the achievability to the third-order approximations by considering shell codes again but with joint maximum-likelihood decoding. Nevertheless, the derivations of these results are based on global error probability formalism and homogeneous blocklength constraints, which are not applicable to the case with heterogeneous error probability and blocklength constraints. Moreover, shell codes and joint decoding are difficult to realize in practice.

In this work, we design a practical multiple access scheme to support heterogeneous services coexistence in the uplink. We restrict the underlying signaling to be discrete signaling formed by binary channel codes and discrete constellations, e.g., quadrature amplitude modulations (QAM), which are the prevailing setup in current communication systems \cite{TS138212_v16p8}. Meanwhile, the receiver employs treating interference as noise (TIN) decoding as it only has single-user decoding complexity and latency. It is worth noting that TIN decoding allows the base station to decode multiple transmission blocks in parallel, such that the decoding of different URLLC services does not need to wait for each other to finish as in SIC decoding. Discrete signaling with TIN with infinite blocklength and homogeneous interference were investigated for the Gaussian broadcast channel (GBC) and the Gaussian interference channel in \cite{8291591} and \cite{9535131}, respectively. Notably, it was proved that under TIN decoding, properly designed discrete signaling can achieve the whole capacity region to within a constant gap for all interference regimes \cite{9535131}. In contrast, when the interference is not very weak, Gaussian signaling with TIN decoding even fails to achieve the optimal generalized degrees of freedom \cite{4675741}, let alone achieving the constant gap. Recently, we proposed a downlink multiple access scheme for the GBC under heterogeneous blocklength and error probability constraints \cite{JSACQiu22}. However, the scheme therein cannot be directly applied to the uplink due to the subtle difference between the broadcast nature (i.e., all users' signals go through the same channel to one receiver) and the multiple access nature (i.e., each user's signal goes through a different channel to the receiver). In this paper, we present new scheme designs and achievable rate analysis for the GMAC with heterogeneous blocklength and error probability constraints, which we term heterogeneous MAC. 
The main contributions are as follows.

\begin{itemize}
\item We approximate the heterogeneous MAC as a cascaded linear deterministic model \cite{Avestimehr11} by capturing the key channel characteristics. The capacity gap (in the asymptotic sense) between each component deterministic MAC and the corresponding component GMAC is bounded within 1 bit/s/Hz. We then identify the achievable rate region and design coding schemes that are proved to achieve the whole rate region with TIN decoding.

    \item  Guided by the deterministic model, we translate the coding schemes therein into the coded modulation schemes for the heterogeneous MAC. Specifically, each user only employs a single channel code to code over multiple sub-blocks of symbol, where the constellation associated with each sub-block can be different depending on the heterogeneity of blocklength and interference strength. The use of point-to-point channel codes and TIN decoding ensures that the encoding and decoding complexities and latencies of each user's messages are the same as in the single-user case.

\item We analyze the mutual information achieved by the proposed coded modulation scheme and rigorously prove that the gap between it and the GMAC capacity is upper bounded by a constant independent of the number of users and channel coefficients. Interestingly, such performance is achieved without the need to use all of the transmit power budget, leading to potential power savings. Further, we characterize the second-order achievable rate of each user by taking into account the heterogeneous blocklength and error probability constraints. Guidelines on how to determine the channel coding and modulation parameters based on the target second-order achievable rate are then provided.

\item Numerical results are provided to demonstrate that the finite blocklength achievable rate of the proposed scheme with QAM and TIN decoding is close to and sometimes even larger than that of Gaussian signaling and perfect SIC decoding. This is due to the close-to-capacity first-order term (i.e., mutual information) and smaller second-order term (i.e., dispersion). To confirm the effectiveness and practicality of the proposed scheme, we further investigate and verify the error probability of the proposed scheme by employing the 5G standard channel coding and modulations.

\end{itemize}

\subsection{Notations}
All logarithms are base 2. Random variables are written in upper case sans-serif fonts, e.g., $\msf{X}$, while their realizations are in lower case form, e.g., $x$. The mean and variance are denoted by $\E[\cdot]$ and $\text{Var}[\cdot]$, respectively. $\msf{X}^{[n]}$ denotes the sequence $\msf{X}[1],\ldots,\msf{X}[n]$. $\msf{X}_1 \overset{d}{=}\msf{X}_2$ means that $\msf{X}_1$ and $\msf{X}_2$ are equal in distribution. $\msf{X}\overset{\text{unif}}{\sim} \text{QAM}(|\Lambda|,d_{\min}(\Lambda))$ represents that $\msf{X}$ is uniformly distributed over a zero mean regular QAM $\Lambda$ with cardinality $|\Lambda|$ and minimum distance $d_{\min}(\Lambda)$, whose average energy is given by $E_{\Lambda} = d_{\min}^2(\Lambda)\frac{|\Lambda|-1}{6}$. $\lceil x\rceil$ rounds $x$ to the nearest integer greater than or equal to $x$. We define the operation $(x)^+ \triangleq \max\{0,x\}$. The binary field, the collections of binary vectors of size $n$ and binary matrices of size $m\times n$ are denoted by $\mathbb{F}_2$, $\mathbb{F}^n_2$, and $\mathbb{F}_2^{m,n}$, respectively. Mutual information, entropy, and differential entropy are denoted by $\mathds{I}(\cdot ; \cdot)$, $\mathbb{H}(\cdot)$, and $\mathbb{H}_\text{d}(\cdot)$, respectively.

\section{System Model}\label{sec:model}
We consider an uplink MAC that consists of $K$ senders and one receiver. Let $k\in \{1,\ldots,K\}$ be the user index. Let $\pmb{x}_k\in\mathbb{C}^{N_k}$ represent the coded symbols of user $k$, where $N_k$ is the symbol length. Due to heterogeneous blocklength constraints, we assume $N_1 \leq\ldots\leq N_K$ without loss of generality. Each user's coded symbols satisfy the individual maximum power constraint per codeword \cite[Eq. (192)]{5452208} given by $\pmb{M}$
\begin{align}\label{eq:p_constraint}
\frac{1}{N_k}\sum^{N_k}_{j=1}|x_k[j]|^2\leq P_k.
\end{align}
For ease of presentation, we further assume that $K$ users transmit their packets to the receiver at the same starting time. It is possible that users can transmit at different starting times. This, however, does not affect the proposed design principle and analysis in the subsequent sections because low-complexity TIN decoding is adopted at the receiver. On the other hand, this could complicate any SIC-based decoding schemes.

\begin{figure}[t!]
	\centering
\includegraphics[width=0.75\linewidth]{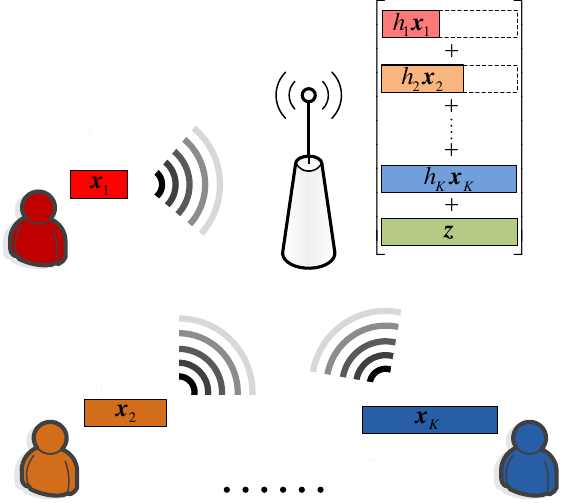}
\caption{$K$ users send their packets with different lengths to the receiver.}
\label{fig:sys1}
\end{figure}

An illustration of the system model is in Fig, \ref{fig:sys1}. Let $h_k\in\mathbb{C}$ represent the complex channel from user $k$ to the receiver. We assume that $h_k$ does not change during $N_k$ symbol periods. At the receiver, a total number of $N_K$ symbols are received. The $j$-th received symbol for $j\in\{1,\ldots,N_K\}$ is
\begin{align}\label{eq:model}
y[j]= \left\{ \begin{array}{l}
\sum^K_{k=1}h_kx_k[j]+z[j],\; j=1,\ldots,N_1\\
\sum^K_{k=k'}h_kx_k[j] + z[j],\; j=N_{k'-1},\ldots,N_{k'}\\
\qquad \qquad \qquad \qquad \quad \;\; k'\in\{2,\ldots,K-1\}\\
h_{K}x_{K}[j]+z[j],j=N_{K-1}+1,\; \ldots,N_K\\
\end{array} \right.,
\end{align}
where $z[j] \sim \mathcal{CN}(0,1)$ is the i.i.d. Gaussian noise for $j\in \{1,\ldots,N_{K}\}$. In addition, we assume the channel state information (CSI) is available so that each transmitter can eliminate the channel phase by rotating its signal by $\frac{h_k^*}{|h_k|}$. In this regard, the channel model in \eqref{eq:model} can be equivalently expressed as that with some $h_k\in \mathbb{R}$.

From Fig. \ref{fig:sys1} and Eq. \eqref{eq:model}, it can be seen that the longer transmission blocks suffer from \emph{heterogeneous interference}. For instance, user $K$'s transmitted symbol $x_K[j]$ is interfered by the other $K-1$ users for $j=1,\ldots,N_1$, interfered by users $2,\ldots,K-1$ for $j=N_1+1,\ldots,N_2$ and so on and so forth. However, user $K$'s symbol $x_K[j]$ is interference free for $j=N_{K-1}+1,\ldots,N_K$. In this regard, it is possible to divide the transmitted symbol block of user $k$ into $k$ sub-blocks as
\begin{align}\label{eq:xk_divide}
\pmb{x}_k = [\pmb{x}_{k,1},\ldots,\pmb{x}_{k,k}],
\end{align}
where for $\ell_k\in\{1,\ldots,k\}$, $\pmb{x}_{k,\ell_k}$ is the $\ell_k$-th sub-block\footnote{Strictly speaking, $\pmb{x}_k$ is decomposed into $\sum^k_{\ell_k=2}\mathds{1}_{\{N_{\ell_k-1} \neq N_{\ell_k}\}}+1 \leq k$ sub-blocks, where $\mathds{1}_{\{.\}}$ is the indicator function. For ease of understanding, we present the case of $N_{\ell_k-1} \neq N_{\ell_k},\forall \ell_k\in\{2,\ldots,k\}$. However, whether this condition holds or not does not affect the proposed scheme.}\footnote{Since user 1's transmitted block has the shortest length, thus $\pmb{x}_{1,1} = \pmb{x}_1$. In the rest of the paper, both subscripts $_{1,1}$ and $_1$ have the same meaning for all the arguments associated with user 1.} interfering with users with index set $\{\ell_k,\ldots,K\}\setminus k$. We make no assumption on the type of users. In other words, all users can be different types of URLLC users or a combination of URLLC and other user types. We denote by $\SNR_k=P_k|h_k|^2$ and $\epsilon_k$ the signal-to-noise ratio (SNR) and the required decoding error probability of user $k$, respectively. Moreover, a rate tuple $(R_1, \ldots, R_K)$ is said to be achievable if there exists an $(N_k,M_k,P_k,\epsilon_k)$ code for user $k$ that can send $M_k$ messages within $N_k$ channel uses with each codeword satisfying the power constraint $P_k$ in \eqref{eq:p_constraint}, $\frac{\log M_k}{N_k} \geq R_k$, and an average probability of error not exceeding $\epsilon_k$, for every $k\in\{1,\ldots,K\}$ simultaneously.

In this work, we consider a coordinated multiple access scenario. Although our model assumes full CSI, additional constraints, namely heterogeneous (finite) transmission blocklength and error probability requirements for each user are taken into account. We note that for this model, many problems, such as the optimal communication strategy and the fundamental limit, remain unsolved. In this regard, we believe that a natural and logical approach would be studying this fundamental model first rather than directly jumping into general channel models without a deep understanding. We leave the extension of the proposed scheme for more general models, e.g., random arrival without assuming perfect synchronization and CSI, in our future work.

\section{The Deterministic Channel and Achievable Schemes with TIN}\label{sec:scheme}
In this section, we first approximate the channel model in Section \ref{sec:model} as a cascaded linear deterministic model \cite{Avestimehr11} by capturing the key channel characteristics. We then construct the coding scheme with TIN decoding to achieve the whole capacity region of the deterministic model. The deterministic model will provide guidance and insight into designing the discrete input distributions for the heterogeneous MAC.

\subsection{Deterministic MAC under Heterogeneous Interference}\label{sec:det_model}
From \eqref{eq:model}, we know that at time $j=N_{k'-1},\ldots,N_{k'}$ and for $k'\in\{1,\ldots,K-1\}$, the channel is a classic GMAC with $K-k'+1$ uplink users transmitting simultaneously. We approximate each component classic GMAC as a deterministic MAC \cite{Avestimehr11} such that the resultant deterministic channel is a concatenation of $K$ component deterministic MACs. Essentially, multiple access is modeled as a deterministic process, which can be seen as multiple links connected to a bit pipe that only passes the bits above the noise level. To this end, let $n_k \triangleq \lceil \log  \SNR_k \rceil^+$ be the deterministic channel condition of user $k\in\{1,\ldots,K\}$, which is also its approximated single-user capacity. Moreover, we define $q_k \triangleq \max\{n_k,\ldots,n_K\}$ to be maximum deterministic channel gain among users $k,\ldots,K$. Note that such an approximation is only used in the deterministic MAC to guide our design. For the proposed achievable scheme for the heterogeneous MAC in Section \ref{sec:cm}, we always work with the actual channel SNR, while $n_k$ is a design parameter. In addition, we consider that user $k$'s transmitted codeword vector can be decomposed into $k$ sub-codewords $\msf{X}_{k,1},\ldots,\msf{X}_{k,k}$ similar to \eqref{eq:xk_divide}. The deterministic channel model corresponding to \eqref{eq:model} is
\begin{align}
\msf{Y}_1=& \bigoplus^K_{k=1}\pmb{S}^{q_1-n_k}\msf{X}_{k,1},\label{eq:det_model1} \\
\msf{Y}_k=& \bigoplus^K_{k'=k}\pmb{S}^{q_{k}-n_{k'}}\msf{X}_{k',k},\; k\in\{2,\ldots,K-1\},\label{eq:det_model2}\\
\msf{Y}_K=&\pmb{S}^{q_K-n_K}\msf{X}_{K,K}, \label{eq:det_model3}
\end{align}
where the multiplication and summation $\oplus$ are over $\mathbb{F}_2$, the $k$-th component deterministic channel input and output are binary column vectors of size $q_k$, i.e., $\msf{X}_{k',k},\msf{Y}_k\in\mathbb{F}^{q_k}_2$ for $k'\in\{k,\ldots,K\}$, and $\pmb{S}$ is a $q_k \times q_k$ down shift matrix. The operational meaning of the binary column vector is that each of its entries represents a power level or a bit. $\pmb{S}^{q_k-n_{k'}}\msf{X}_{k',k}$ models the channel effect, where the lowest $q_k-n_{k'}$ bits of $\msf{X}_{k',k}$ are shifted down below the noise level and truncated.

\subsection{The Two-User Case}
\begin{figure}[t!]
	\centering
\includegraphics[width=0.55\linewidth]{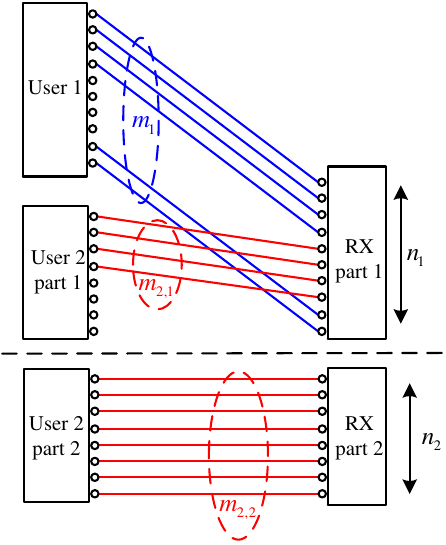}
\caption{The two-user deterministic model with $(n_1,n_2)=(10,8)$ and $(m_1,m_{2,1},m_{2,2})=(6,4,8)$.}
\label{fig:sys2}
\end{figure}

For ease of understanding, we first present the two-user case with $K=2$, identify the achievable rate region and propose an achievable scheme. We assume $n_1>n_2$ and let $q\triangleq \max\{n_1,n_2\}=n_1$. Recall that user 2's codeword can be decomposed into two parts. Let $m_1$ be the number of successfully transmitted bits or achievable rate for user 1, $m_{2,1}$ and $m_{2,2}$ be the numbers of successfully transmitted bits of user 2 for the first and second parts, respectively. For the channel model in \eqref{eq:det_model1}-\eqref{eq:det_model3} with $K=2$, the set of non-negative rate tuples $(m_1,m_{2,1},m_{2,2})$ are achievable if
\begin{align}
m_1+m_{2,1} \leq& n_1, \label{eq:cap1}\\
m_{2,1} \leq& n_{2},\label{eq:cap2}\\
m_{2,2} \leq & n_{2}.\label{eq:cap3}
\end{align}
Here, two inequalities \eqref{eq:cap1} and \eqref{eq:cap2} together give the two-user deterministic MAC capacity region while \eqref{eq:cap3} is the deterministic point-to-point channel capacity. Both deterministic capacities are \emph{within} 1 bit/s/Hz to the capacities of their Gaussian counterparts \cite{Avestimehr11}, respectively. An illustration for this model is given in Fig. \ref{fig:sys2}. It is worth emphasizing that the considered deterministic MAC is different from the one in \cite{Avestimehr11}, where user 2's bits all experience the same level of interference strength. More importantly, our goal is to design the input distributions $(\msf{X}_1,\msf{X}_{2,1},\msf{X}_{2,2})$ to achieve the above rate region with TIN decoding only.

Let $\msf{U}_1\in\mathbb{F}^{m_1}_2$ and $\msf{U}_{2,1}\in\mathbb{F}^{m_{2,1}}_2$ be users 1 and 2's message vectors, respectively, with each entry drawn independently and uniformly distributed over $\mathbb{F}_2$. Let $\pmb{G}_1 \in \mathbb{F}_2^{q,m_1}$ and $\pmb{G}_{2,1} \in \mathbb{F}_2^{q,m_{2,1}}$ be the generator matrices such that $\msf{X}_1 = \pmb{G}_1 \msf{U}_1$ and $\msf{X}_{2,1} = \pmb{G}_{2,1} \msf{U}_{2,1}$. The achievable rate of user 1 under TIN can be derived as
\begin{subequations}\label{eq:IX1Y1}
\begin{align}
&\mathds{I}(\msf{X}_1;\msf{Y}_1) = \mathbb{H}(\msf{Y}_1) - \mathbb{H}(\msf{Y}_1|\msf{X}_1) \label{eq:I10}  \\
 =& \mathbb{H}(\pmb{S}^{q-n_1}\pmb{G}_1 \msf{U}_1\oplus \pmb{S}^{q-n_{2}}\pmb{G}_{2,1} \msf{U}_{2,1}) \nonumber\\
 &- \mathbb{H}(\pmb{S}^{q-n_{2}}\pmb{G}_{2,1} \msf{U}_{2,1})   \\
 =& \text{rank}([\pmb{S}^{q-n_{k}}\pmb{G}_1,\pmb{S}^{q-n_{2}}\pmb{G}_{2,1}])
 - \text{rank}(\pmb{S}^{q-n_{2}}\pmb{G}_{2,1}), \label{eq:I1}
\end{align}
\end{subequations}
where the multiplication and addition $\oplus$ are over $\mathbb{F}_2$. User 2's rate under TIN $\mathds{I}(\msf{X}_{2,1};\msf{Y}_1)$ can be easily obtained from \eqref{eq:IX1Y1} by swapping the arguments between subscripts ``1'' and ``2''. From this point onward, our goal is to design the generator matrices $\pmb{G}_1$, $\pmb{G}_{2,1}$, and $\pmb{G}_{2,2}$ to achieve the rate region of \eqref{eq:cap1}-\eqref{eq:cap3}.

First, to achieve the rate region of \eqref{eq:cap1}-\eqref{eq:cap2}, we propose
\begin{align}\label{eq:G1}
\pmb{G}_1=\begin{bmatrix}
\pmb{0}^{n_{1}-m_1-m_{2,1},m_1}\\
\pmb{F}_{1}\\
\pmb{0}^{m_{2,1},m_1}\\
\end{bmatrix},
\end{align}
and
\begin{align}\label{eq:G2}
\pmb{G}_{2,1}=\begin{bmatrix}
\pmb{0}^{n_{2}-m_{2,1},m_{2,1}}\\
\pmb{F}_{2,1}\\
\pmb{0}^{n_1-n_{2},m_{2,1}}\\
\end{bmatrix},
\end{align}
where $\pmb{F}_1\in\mathbb{F}^{m_1,m_1}_2$ and $\pmb{F}_{2,1}\in\mathbb{F}^{m_{2,1},m_{2,1}}_2$ are submatrices with linearly independent rows. With the proposed $\pmb{G}_1$ and $\pmb{G}_{2,1}$, we have that
\begin{subequations}\label{eq:1_term}
\begin{align}
&\text{rank}([\pmb{S}^{q-n_{1}}\pmb{G}_1,\pmb{S}^{q-n_{2}}\pmb{G}_{2,1}])  \nonumber \\
=&\text{rank}\left(
\begin{bmatrix}
\left. {\begin{array}{*{20}{c}}
\pmb{0}^{n_{1}-m_1-m_{2,1},m_1}\\
\pmb{F}_{1}\\
\end{array}} \right\} & \left\{\begin{array}{*{20}{c}} \pmb{0}^{n_1-n_{2},m_{2,1}}\\ \pmb{0}^{n_{2}-m_{2,1},m_{2,1}}\\\end{array} \right. \\
\pmb{0}^{m_{2,1},m_1} & \pmb{F}_{2,1}\\
\end{bmatrix}
\right)\\
=& m_1+m_{2,1},
\end{align}
\end{subequations}
and
\begin{align}
\text{rank}(\pmb{S}^{q-n_{2}}\pmb{G}_{2,1})=\text{rank}(\pmb{F}_{2,1}) =m_{2,1}. \label{eq:2_term}
\end{align}
Substituting \eqref{eq:1_term} and \eqref{eq:2_term} into \eqref{eq:IX1Y1} results in $\mathds{I}(\msf{X}_1;\msf{Y}_1)=m_1$. Similarly, we can follow the above steps to obtain that $\mathds{I}(\msf{X}_{2,1};\msf{Y}_1)=m_{2,1}$.

Let $\msf{U}_{2,2}\in\mathbb{F}^{m_{2,2}}_2$ be another part of user 2's message vector. To achieve the rate in \eqref{eq:cap3}, we design
\begin{align}\label{eq:G22}
\pmb{G}_{2,2}=\begin{bmatrix}
\pmb{F}_{2,2}\\
\pmb{0}^{n_{2}-m_{2,2},m_{2,2}}\\
\end{bmatrix},
\end{align}
where $\pmb{F}_{2,2}\in\mathbb{F}^{m_{2,2},m_{2,2}}_2$. Then, we have that $\msf{X}_{2,2} = \pmb{G}_{2,2} \msf{U}_{2,2}$, which is interference-free according to \eqref{eq:det_model3} with $K=2$. As a result, we have that $\mathds{I}(\msf{X}_{2,2};\msf{Y}_2)=\text{rank}(\pmb{F}_{2,2})=m_{2,2}$. From here, we see that the proposed scheme can achieve the rate region of \eqref{eq:cap1}-\eqref{eq:cap3} with TIN decoding.

When $n_1<n_2$, the design of $\pmb{G}_1$ and $\pmb{G}_{2,1}$ can be easily obtained from \eqref{eq:G1} and \eqref{eq:G2} by swapping the argument between subscripts ``1'' and ``2''. The design of $\pmb{G}_{2,2}$ remains unchanged.

\subsection{The $K$-User Case}\label{sec:det_k_user}
In this section, we propose achievable schemes for the $K$-user case. We first consider the case $\SNR_1>\ldots>\SNR_K$ such that $n_1>\ldots>n_K$ and $q_k \triangleq \max\{n_k,\ldots,n_K\}=n_k$. For $\ell_k\in\{1,\ldots,k\}$, let $m_{k,\ell_k}$ be the number of bits that user $k$ successfully transmits at its $\ell_k$-th component channel. For the channel model in \eqref{eq:det_model1}-\eqref{eq:det_model3}, the achievable rate region $(m_{1},m_{2,1},\ldots,m_{K,K})$ is the union of those of $K$ component deterministic MAC channels, i.e.,
\begin{align}\label{eq:rate_region_dK}
\sum^K_{k'=k}m_{k',k} \leq n_k, \; k\in\{1,\ldots,K\}.
\end{align}

For user $k$, let $\msf{U}_{k,\ell_k}\in\mathbb{F}^{m_{k,\ell_k}}_2$ be its $\ell_k$-th component message vector, where $\ell_k\in\{1,\ldots,k\}$. Let $\pmb{G}_{k,\ell_k} \in \mathbb{F}_2^{q_{\ell_k},m_{k,\ell_k}}$ be the generator matrix such that the $\ell_k$-th component codeword satisfies $\msf{X}_{k,\ell_k} = \pmb{G}_{k,\ell_k}\msf{U}_{k,\ell_k}$, and $q_{\ell_k}=n_{\ell_k}$. For the $\ell_k$-th component deterministic MAC of \eqref{eq:det_model1}-\eqref{eq:det_model3}, the codewords of user $k$ and the users of index group $\{\ell_k,\ell_k+1,\ldots,K\}\setminus k$ interfere with each other. The achievable rate of user $k$ under TIN decoding can be derived as
\begin{subequations}\label{eq:I_XkYk}
\begin{align}
&\mathds{I}(\msf{X}_{k,\ell_k};\msf{Y}_{\ell_k}) = \mathbb{H}(\msf{Y}_{\ell_k}) - \mathbb{H}(\msf{Y}_{\ell_k}|\msf{X}_{k,\ell_k})   \\
 =& \mathbb{H}\left(\bigoplus^K_{k'=\ell_k}\pmb{S}^{q_{\ell_k}-n_{k'}}\pmb{G}_{k',\ell_k} \msf{U}_{k',\ell_k}\right) \nonumber\\
 &- \mathbb{H}\left(\bigoplus_{k'\in\{\ell_k,\ldots,K\}\setminus k}\pmb{S}^{q_{\ell_k}-n_{k'}}\pmb{G}_{k',\ell_k} \msf{U}_{k',\ell_k}\right)   \\
 =&\text{rank}([\pmb{S}^{q_{\ell_k}-n_{\ell_k}}\pmb{G}_{\ell_k,\ell_k},\ldots,\pmb{S}^{q_{\ell_k}-n_{K}}\pmb{G}_{K,\ell_k}]) \nonumber \\
 &- \text{rank}([\pmb{S}^{q_{\ell_k}-n_{\ell_k}}\pmb{G}_{\ell_k,\ell_k},\ldots,\pmb{S}^{q_{\ell_k}-n_{k-1}}\pmb{G}_{k-1,\ell_k},\nonumber \\
 &\pmb{S}^{q_{\ell_k}-n_{k+1}}\pmb{G}_{k+1,\ell_k},\ldots,\pmb{S}^{q_{\ell_k}-n_{K}}\pmb{G}_{K,\ell_k}]), \label{eq:I_kk}
\end{align}
\end{subequations}
where the multiplication and addition $\oplus$ are over $\mathbb{F}_2$. From \eqref{eq:I_XkYk}, our goal becomes designing generator matrices $\pmb{G}_{\ell_k,\ell_k},\pmb{G}_{\ell_k+1,\ell_k},\ldots,\pmb{G}_{K,\ell_k}$ to achieve the target rate region. 

As above, we first focus on the $\ell_k$-th component deterministic MAC. For user $k$' with $k'\in\{\ell_k,\ldots,K\}$, we propose its $\ell_k$-th component generator matrix $\pmb{G}_{k',\ell_k}$ as
\begin{align}\label{Eq:GK}
\pmb{G}_{k',\ell_k}=\begin{bmatrix}
\pmb{0}^{n_{k'}-\sum^{K}_{i=k'}m_{i,\ell_k},m_{k',\ell_k}}\\
\pmb{F}_{k',\ell_k}\\
\pmb{0}^{\sum^{K}_{i=k'+1}m_{i,\ell_k},m_{k',\ell_k}}\\
\pmb{0}^{n_{\ell_k}-n_{k'},m_{k',\ell_k}}\\
\end{bmatrix},
\end{align}
where $\pmb{F}_{k',\ell_k}\in\mathbb{F}^{m_{k',\ell_k},m_{k',\ell_k}}_2$.

Combining the generator matrix with the deterministic channel effect, we have
\begin{subequations}\label{eq:sgk_p}
\begin{align}
\pmb{S}^{q_{\ell_k}-n_{k'}}\pmb{G}_{k',\ell_k}=&
\begin{bmatrix}
\pmb{0}^{n_{\ell_k}-n_{k'},m_{k',\ell_k}}\\
\pmb{0}^{n_{k'}-\sum^{K}_{i=k'}m_{i,\ell_k},m_{k',\ell_k}}\\
\pmb{F}_{k',\ell_k}\\
\pmb{0}^{\sum^{K}_{i=k'+1}m_{i,\ell_k},m_{k',\ell_k}}\\
\end{bmatrix} \label{eq:sgk_p0}\\
=&\begin{bmatrix}
\pmb{0}^{n_{\ell_k}-n_{k},m_{k',\ell_k}}\\
\pmb{0}^{n_{k}-n_{k'},m_{k',\ell_k}}\\
\pmb{0}^{n_{k'}-\sum^{K}_{i=k'}m_{i,\ell_k},m_{k',\ell_k}}\\
\pmb{F}_{k',\ell_k}\\
\pmb{0}^{\sum^{K}_{i=k'+1}m_{i,\ell_k},m_{k',\ell_k}}\\
\end{bmatrix} \label{eq:sgk_p1}\\
=&
\begin{bmatrix}
\pmb{0}^{n_{\ell_k}-n_{k},m_{k',\ell_k}}\\
\pmb{0}^{n_{k}-\sum^{K}_{i=k+1}m_{i,\ell_k},m_{k',\ell_k}}\\
\pmb{0}^{\sum^{k'-1}_{i=k+1}m_{i,\ell_k},m_{k',\ell_k}}\\
\pmb{F}_{k',\ell_k}\\
\pmb{0}^{\sum^{K}_{i=k'+1}m_{i,\ell_k},m_{k',\ell_k}}
\end{bmatrix},
\end{align}
\end{subequations}
where \eqref{eq:sgk_p1} holds for any $k<k'$ and $k\in\{\ell_k,\ldots,K\}$. From \eqref{eq:sgk_p}, we also obtain that
\begin{align}\label{eq:rank_kp}
\text{rank}(\pmb{S}^{q_{\ell_k}-n_{k'}}\pmb{G}_{k',\ell_k})=m_{k',\ell_k}.
\end{align}

Next, we want to show that the following holds by using the proposed generator matrices
\begin{align}\label{eq:rank_sum_k_kp}
\text{rank}&([\pmb{S}^{q_{\ell_k}-n_{k}}\pmb{G}_{k,\ell_k},\pmb{S}^{q_{\ell_k}-n_{k'}}\pmb{G}_{k',\ell_k}])\nonumber \\
=&\text{rank}(\pmb{S}^{q_{\ell_k}-n_{k}}\pmb{G}_{k,\ell_k})
+\text{rank}(\pmb{S}^{q_{\ell_k}-n_{k'}}\pmb{G}_{k',\ell_k}).
\end{align}
To do so, we first substitute $k'=k$ into \eqref{Eq:GK} to obtain $\pmb{G}_{k,\ell_k}$. Then, we can compute the rank of $[\pmb{S}^{q_{\ell_k}-n_{k}}\pmb{G}_{k,\ell_k},\pmb{S}^{q_{\ell_k}-n_{k'}}\pmb{G}_{k',\ell_k}]$ as shown on the top of page 6. As a result, \eqref{eq:rank_sum_k_kp} is obtained from the last equality of \eqref{eq:rank_k_sum}.
\begin{figure*}[t]
\begin{subequations}\label{eq:rank_k_sum}
\begin{align}
\text{rank}([\pmb{S}^{q_{\ell_k}-n_{k}}\pmb{G}_{k,\ell_k},\pmb{S}^{q_{\ell_k}-n_{k'}}\pmb{G}_{k',\ell_k}])  
&\overset{\eqref{eq:sgk_p}}{=}\text{rank}\left(
\begin{bmatrix}
\pmb{0}^{n_{\ell_k}-n_{k},m_{k,\ell_k}} & \pmb{0}^{n_{\ell_k}-n_{k},m_{k',\ell_k}}\\
\left.\begin{array}{*{20}{c}}
\pmb{0}^{n_{k}-\sum^{K}_{i=k}m_{i,\ell_k},m_{k,\ell_k}}\\
\pmb{F}_{k,\ell_k}\\
\end{array}\right\} & \pmb{0}^{n_{k}-\sum^{K}_{i=k+1}m_{i,\ell_k},m_{k',\ell_k}}\\
\pmb{0}^{\sum^{K}_{i=k+1}m_{i,\ell_k},m_{k,\ell_k}}& \left\{\begin{array}{*{20}{c}}
\pmb{0}^{\sum^{k'-1}_{i=k+1}m_{i,\ell_k},m_{k',\ell_k}}\\
\pmb{F}_{k',\ell_k}\\
\pmb{0}^{\sum^{K}_{i=k'+1}m_{i,\ell_k},m_{k',\ell_k}}
\end{array} \right. \\
\end{bmatrix}
\right)\\
&=m_{k,\ell_k}+m_{k',\ell_k}\\
&\overset{\eqref{eq:rank_kp}}{=}\text{rank}(\pmb{S}^{q_{\ell_k}-n_{k}}\pmb{G}_{k,\ell_k})+\text{rank}(\pmb{S}^{q_{\ell_k}-n_{k'}}\pmb{G}_{k',\ell_k}).
\end{align}
\end{subequations}
\hrule
\end{figure*}
It should be noted that \eqref{eq:rank_sum_k_kp} holds under the proposed design, which may not hold in general. Since \eqref{eq:rank_sum_k_kp} holds for any $k<k'$ and $k,k'\in\{\ell_k,\ldots,K\}$, one can also obtain that
\begin{subequations}\label{eq:rank_k_sum1}
\begin{align}
\text{rank}&([\pmb{S}^{q_{\ell_k}-n_{\ell_k}}\pmb{G}_{\ell_k,\ell_k},\ldots,\pmb{S}^{q_{\ell_k}-n_{K}}\pmb{G}_{K,\ell_k}])\nonumber \\
=&\sum^K_{k'=\ell_k}\text{rank}(\pmb{S}^{q_{\ell_k}-n_{k'}}\pmb{G}_{k',\ell_k})\\
=&\sum^K_{k'=\ell_k} m_{k',\ell_k}.
\end{align}
\end{subequations}
Substituting \eqref{eq:rank_k_sum1} into \eqref{eq:I_XkYk} gives
\begin{subequations}
\begin{align}
\mathds{I}(\msf{X}_{k,\ell_k};\msf{Y}_{\ell_k})
 &=\sum^K_{k'=\ell_k} m_{k',\ell_k} - \sum_{k'\in\{\ell_k,\ldots,K\}\setminus k} m_{k',\ell_k} \\
 &=m_{k,\ell_k}.
 \end{align}
 \end{subequations}
As a result, we see that the rate region in \eqref{eq:rate_region_dK} is achievable with TIN decoding because
\begin{align}\label{eq:Ik_region}
\sum^K_{k'=\ell_k}\mathds{I}(\msf{X}_{k',\ell_k};\msf{Y}_{\ell_k})  \leq n_{\ell_k},
\end{align}
for $\ell_k \in\{1,\ldots,k\}$.

Motivated by the capacity-achieving property of the above deterministic scheme and the small gap between the component deterministic and Gaussian model, we will translate the proposed coding scheme for the deterministic channel into capacity-approaching coded modulation scheme for the heterogeneous MAC model in Section \ref{sec:cm}.

\begin{remark}\label{remark:scheme}
The achievable scheme for the deterministic channel is not unique. For example, it can be shown that the following generator matrices
\begin{align}
\pmb{G}_{k',\ell_k}=\begin{bmatrix}
\pmb{F}_{k',\ell_k,a}\\
\pmb{0}^{\sum^{K}_{i=k'+1}m_{i,\ell_k},m_{k',\ell_k}}\\
\pmb{F}_{k',\ell_k,b}\\
\pmb{0}^{n_{k'}-\sum^{K}_{i=\ell_k}m_{i,\ell_k},m_{k',\ell_k}}\\
\pmb{0}^{n_{\ell_k}-n_{k'},m_{k',\ell_k}}\\
\end{bmatrix},\label{Eq:GKa}
\end{align}
for $k'<K$ and $\pmb{F}_{k',\ell_k,a}\in \mathbb{F}^{n_{k'}-n_{k'+1},m_{k',\ell_k}}_2$ and $\pmb{F}_{k',\ell_k,b}\in \mathbb{F}^{(m_{k',\ell_k}-(n_{k'}-n_{k'+1}))^+,m_{k',\ell_k}}_2$, and
\begin{align}
\pmb{G}_{K,\ell_k}=\begin{bmatrix}
\pmb{F}_{k,\ell_k}\\
\pmb{0}^{n_K-m_{K,\ell_k},m_{K,\ell_k}}\\
\pmb{0}^{n_{\ell_k}-n_{K},m_{K,\ell_k}}\\
\end{bmatrix},\label{Eq:GKb}
\end{align}
for $k'=K$ and $\pmb{F}_{K,\ell_k}\in \mathbb{F}^{m_{K,\ell_k},m_{K,\ell_k}}_2$, also lead to a capacity-achieving scheme. Note that this is the generalization of the two-user achievable scheme in our conference version \cite{Qiu23}. The achievability proof directly follows from \eqref{eq:sgk_p0}-\eqref{eq:Ik_region}. Although different generator matrices lead to the same achievable rate region in the deterministic channel, the coded modulation schemes built from these deterministic achievable schemes could have different rates.  \demo
\end{remark}

The proposed scheme can be easily adapted to other channel ordering. For instance, consider $\SNR_{\pi(1)}>\ldots >\SNR_{\pi(K)}$ for some permutation function $\pi(.)$ with permutation length $K$. Based on this, a sub-ordering for the $\ell_k$-th component deterministic MAC can be written as $\SNR_{\pi'(\pi(1))}>\ldots>\SNR_{\pi'(\pi(K-\ell_k+1))}$ for some permutation function $\pi'(.)$ with permutation length $K-\ell_k+1$. In this case, the last entry of the generator matrix in \eqref{Eq:GK} is modified to $\pmb{0}^{n_{\pi'(\pi(1))}-n_{k'},m_{k',\ell_k}}$. This is because the achievable scheme based on \eqref{Eq:GK} for user $k'$ only requires the CSI of its own, i.e., $n_{k'}$, and the user with the largest channel gain among user group $\{\ell_k,\ldots,K\}$, i.e., $n_{\pi'(\pi(1))}$.

\section{Proposed Coded Modulation Scheme With TIN}\label{sec:cm}
In this section, we translate the proposed achievable scheme in Section \ref{sec:det_model} into the coded modulation scheme for the heterogeneous MAC. The performance of the proposed scheme will then be rigorously analyzed. To begin, we assume that $\SNR_1>\ldots>\SNR_K$.

\subsection{Proposed Achievable Scheme}\label{sec:QAM_CM}
We adopt the same channel ordering assumption assumption as in Section \ref{sec:det_k_user} for illustrative purposes. The proposed scheme consists of the following steps.

\subsubsection{Encoding}
User $k$ encodes its length-$I_k$ message $\pmb{u}_k$ into a length-$L_k$ binary codeword $\pmb{c}_k$. Then, codeword $\pmb{c}_k$ is interleaved by employing the bit-interleaved coded modulation (BICM) technique \cite{CIT-019}, which is a flexible and capacity-approaching way for mapping binary coded bits onto high-order modulation symbols. The interleaved codeword $\tilde{\pmb{c}}_k$ is mapped to a length-$N_k$ symbol sequence $\pmb{x}_k$, where the modulation design and mapping steps will be described next. It is worth emphasizing that each user only uses a \emph{single} channel code such that the encoding complexity is the same as that of the single-user case.

\subsubsection{Constellation Symbol Mapping}\label{sec:2u_const_map}
From \eqref{eq:model}, we note that for $k>1$, user $k$'s transmission symbol block suffers from heterogeneous interference across its symbols. To handle heterogeneous interference, we divide the user $k$'s symbol block into $k$ sub-blocks as in \eqref{eq:xk_divide}, where the $\ell_k$-th sub-block for $\ell_k\in\{1,\ldots,k\}$ contains $N_{\ell_k}-N_{\ell_k-1}$ symbols
\begin{align}
\pmb{x}_{k,\ell_k} = \left[x_k[N_{\ell_k-1}+1],\ldots,x_k[N_{\ell_k}]\right].
\end{align}
One of the key features of the proposed scheme is to allow user $k$ to employ $k$ sets of constellations $\Lambda_{k,1},\ldots,\Lambda_{k,k}$, some of which can be the same or different. The constellation designs will be introduced in Section \ref{sec:2u_const}. For the $\ell_k$-th sub-block of $\pmb{x}_k$, we have that
\begin{align}
x_k[j] \in \Lambda_{k,\ell_k}, \; j= N_{\ell_k-1}+1,\ldots, N_{\ell_k}.
\end{align}
Let $m_{k,\ell_k} = \log|\Lambda_{k,\ell_k}|$ be the modulation order of the $\ell_k$-th modulation set.
User $k$ maps its interleaved codeword $\tilde{\pmb{c}}_k$ to modulated symbol block $\pmb{x}_k$ following the proposed bit mapping rule below
\begin{align}
\tilde{\pmb{c}}_{k,\ell_k} \rightarrow \pmb{x}_{k,\ell_k},\; \ell_k = 1,\ldots,k,
\end{align}
where
\begin{align}
\tilde{\pmb{c}}_{k,\ell_k} =& \bigg[\tilde{c}\left[\sum\nolimits^{k-1}_{\ell_k=1}(N_{\ell_k}-N_{\ell_k-1})m_{k,\ell_k}\right],\ldots,\nonumber \\
&\tilde{c}\left[\sum\nolimits^k_{\ell_k=1}(N_{\ell_k}-N_{\ell_k-1})m_{k,\ell_k}\right] \bigg],
\end{align}
is the $\ell_k$-th sub-codeword of $\tilde{\pmb{c}}_k$. As a result, user $k$'s codeword length satisfies
\begin{align}\label{eq:Lk_Nk}
L_k=\sum^k_{\ell_k=1}(N_{\ell_k}-N_{\ell_k-1})m_{k,\ell_k}.
\end{align}

According to \eqref{eq:model}, user 1's transmission block suffers from homogeneous interference. Hence, user 1 only needs to employ one constellation set $\Lambda_1$ and the bit mapping is the same as in the single-user case, i.e., $\tilde{\pmb{c}}_1 \rightarrow \pmb{x}_1$.

\subsubsection{Constellation Design}\label{sec:2u_const}

\begin{figure*}[t!]
	\centering
\includegraphics[width=0.6\linewidth]{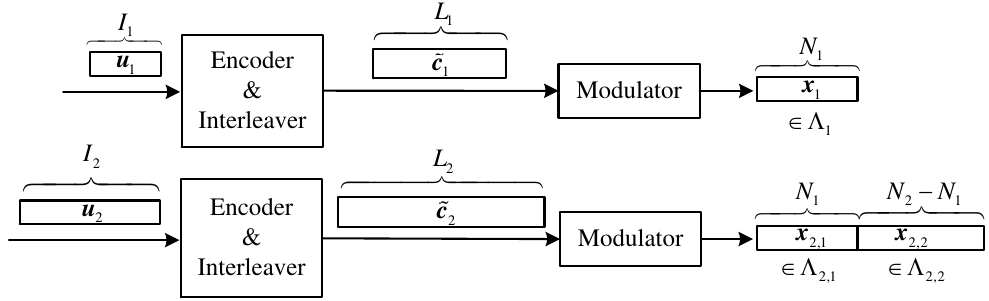}
\caption{Two users send their coded symbols with different lengths to the receiver.}
\label{fig:Henc1}
\end{figure*}

The achievable scheme based on the generator matrix \eqref{Eq:GK} for the deterministic channel is systematically translated into QAM signalings for the Gaussian channel. For user $k\in\{1,\ldots,K\}$, the $j$-th modulation symbol in its $\ell_k$-th sub-block, $\ell_k\in\{1,\ldots,k\}$, is given by
\begin{align}\label{eq:Xk_QAM}
\msf{X}_k[j]&=\eta_{\ell_k} \sqrt{P_k}2^{\frac{n_{\ell_k}-\log\SNR_k+\sum^K_{i=k+1}m_{i,\ell_k}}{2}}\msf{F}_{k,\ell_k}, \\
&\overset{\text{unif}}{\sim}\Lambda_{k,\ell_k},\; \text{for}\;j= N_{\ell_k-1}+1,\ldots, N_{\ell_k},
\end{align}
where $\msf{F}_{k,\ell_k}\overset{\text{unif}}{\sim} \text{QAM}(2^{m_{k,\ell_k}},1)$ with the modulation order $m_{k,\ell_k}$ determined by the rank number of submatrix $\pmb{F}_{k,\ell_k}$ in $\pmb{G}_{k,\ell_k}$ in \eqref{Eq:GK}, $\Lambda_{k,\ell_k}$ is a scaled regular QAM constellation with total cardinality $2^{m_{k,\ell_k}}$, $\eta_{\ell_k}$ is the normalization factor to ensure that $\E[| \msf{X}_k[j]|^2]\leq P_k$ for $j=1,\ldots,N_k$.

Here, we point out that the power coefficient in front of the random variable $\msf{F}_{k,\ell_k}$ is determined by the number of rows below its corresponding submatrix $\pmb{F}_{k,\ell_k}$ in $\pmb{G}_{k,\ell_k}$. In \eqref{Eq:GK}, we see that the number of rows below $\pmb{F}_{k,\ell_k}$ is $n_{\ell_k}-n_k+\sum^K_{i=k+1}m_{i,\ell_k}$, In \eqref{eq:Xk_QAM}, we adopt this row number to form the power coefficient, except that we approximate $\log\SNR_k\approx n_k$. The use of $\log\SNR_k$ ensures that the superimposed constellation has a constant minimum distance lower bound even in the worst case, which will be discussed in Section \ref{sec:analysis}. In addition, the normalization factor for user $k$'s symbol in the $\ell_k$-th sub-block is
\begin{align}\label{eq:eta_nk2}
\eta_{\ell_k} =& \sqrt{\frac{1}{\max\{E_{\ell_k,\ell_k},\ldots,E_{K,\ell_k}\}}},
\end{align}
where we define the average energy of user $k$'s symbol in its $\ell_k$-th sub-block for $k\in\{\ell_k,\ldots,K\}$ as
\begin{align}
E_{k,\ell_k} \triangleq \E\left[\left|2^{\frac{n_{\ell_k}-\log\SNR_k+\sum^K_{i=k+1}m_{i,\ell_k}}{2}}\msf{F}_{k,\ell_k}\right|^2\right].
\end{align}

Finally, user $k$ sends its symbol block according to \eqref{eq:Xk_QAM} to the receiver. An example of the proposed coded modulation scheme in the two-user case is illustrated in Fig. \ref{fig:Henc1}. The achievable scheme in Remark \ref{remark:scheme} can also be translated into a coded modulation scheme in the same manner as to \eqref{eq:Xk_QAM}. Due to space limitations, we do not repeat the same description.

\subsubsection{TIN Decoding}
\begin{figure}[t!]
	\centering
\includegraphics[width=0.9\linewidth]{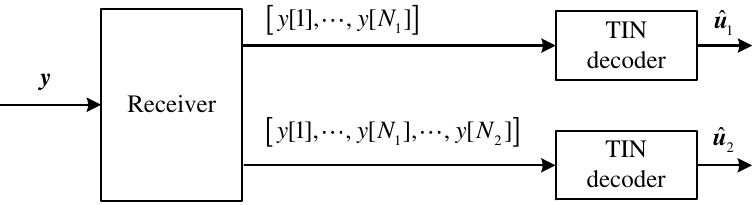}
\caption{The receiver performs parallel TIN decoding on two users' transmitted blocks.}
\label{fig:Hdec1}
\end{figure}

At the receiver, each user's message is decoded by treating other user's signals as noise. Specifically, the receiver starts to decode user 1's message upon receiving $[y[1],\ldots,y[N_1]]$ while the decoding of user 2 starts upon receiving $[y[1],\ldots,y[N_{2}]]$. Similarly, the decoding of user $k$'s message starts one $[y[1],\ldots,y[N_k]]$ is received. It is worth emphasizing that TIN decoding can be performed in parallel as illustrated in Fig. \ref{fig:Hdec1}, which is more favorable if several users are URLLC users. This is in sharp contrast to SIC decoding which requires the reception of the whole transmission block $[y[1],\ldots,y[N_K]]$ and needs to be performed in sequential. The decoding process is the same as that in the point-to-point channel using BICM \cite{CIT-019}. An error occurs at the decoding of user $k$'s message if the decoder's output $\pmb{\hat{u}}_k \neq \pmb{u}_k$.

\subsection{Mutual Information and Minimum Distance Analysis}\label{sec:analysis}
To theoretically demonstrate the effectiveness of the proposed scheme, we analyze the mutual information in the heterogeneous MAC. Note that mutual information plays a crucial role in the finite blocklength achievable rate. To further provide the lower bound on the achievable mutual information, we then analyze the minimum constellation distance at the receiver. Our results show that the mutual information achieved by the proposed scheme is close to capacity.

For user $k\in\{1,\ldots,K\}$, we note that its transmitted symbols $\msf{X}_k[j]$ in its $\ell_k$-th sub-block, $\ell_k\in\{1,\ldots,k\}$, are independent and identically distributed (i.i.d.) for $j=N_{\ell_k-1}+1,\ldots,N_{\ell_k}$, and satisfy $\msf{X}_k[j]\overset{\text{unif}}{\sim}\Lambda_{k,\ell_k}$, where we set $N_0=0$. Consequently, the received symbols $\msf{Y}[j]$ in \eqref{eq:model} are also i.i.d. for $j=N_{k-1}+1,\ldots,N_{k}$. For notation simplicity, we drop the symbol index $[j]$ and let $\msf{X}_{k,\ell_k}\overset{d}{=}\msf{X}_{k}[j]$ for $j=N_{\ell_k-1}+1,\ldots,N_{\ell_k}$ and $\msf{Y}_{k}\overset{d}{=} \msf{Y}[j]$ for $j=N_{k-1}+1,\ldots,N_{k}$.

We first derive the mutual information between $\msf{X}_{k,\ell_k}$ and $\msf{Y}_{\ell_k}$ when user $k$'s whole transmission block is decoded using TIN as
\begin{subequations}\label{eq:Ik_mutual_inf}
\begin{align}
&\mathds{I}(\msf{X}_{k,\ell_k};\msf{Y}_{\ell_k})=\mathbb{H}_\text{d}(\msf{Y}_{\ell_k})-\mathbb{H}_\text{d}(\msf{Y}_{\ell_k}|\msf{X}_{k,\ell_k})  \label{eq:Ik_mutual_inf_step1}\\
=&\mathbb{H}_\text{d}(\msf{Y}_{\ell_k})-\mathbb{H}_\text{d}(\msf{Z})-\mathbb{H}_\text{d}\left(\sum_{i\in\{\ell_k,\ldots,K\} \setminus k }h_{i}\msf{X}_{i,\ell_k}+\msf{Z}\right) \nonumber \\
&+\mathbb{H}_\text{d}(\msf{Z}) \\
=&\mathds{I}\left(\sum^K_{i=\ell_k}h_{i}\msf{X}_{i,\ell_k};\msf{Y}_{\ell_k}\right)\nonumber \\
&-\mathds{I}\left(\sum_{i\in\{\ell_k,\ldots,K\} \setminus k }h_{i}\msf{X}_{i,\ell_k};\sum_{i\in\{\ell_k,\ldots,K\} \setminus k }h_{i}\msf{X}_{i,\ell_k}+\msf{Z}\right) \label{eq:Ik_mutual_inf_step3}\\
\geq&\mathds{I}\left(\sum^K_{i=\ell_k}h_{i}\msf{X}_{i,\ell_k};\msf{Y}_{\ell_k}\right)-\mathbb{H}\left(\sum_{i\in\{\ell_k,\ldots,K\} \setminus k }h_{i}\msf{X}_{i,\ell_k}\right) \\
=& \mathds{I}\left(\sum^K_{i=\ell_k}h_{i}\msf{X}_{i,\ell_k};\msf{Y}_{\ell_k}\right)- \sum_{i\in\{\ell_k,\ldots,K\} \setminus k }m_{i,\ell_k}. \label{eq:Ik_first_term}
\end{align}
\end{subequations}
It remains to bound the first term in \eqref{eq:Ik_first_term}, i.e., $\mathds{I}(\sum^K_{i=\ell_k}h_{i}\msf{X}_{i,\ell_k};\msf{Y}_{\ell_k})$. To do so, we first analyze the minimum Euclidean distance of the superimposed constellation $\sum^K_{i=\ell_k}h_{i}\msf{X}_{i,\ell_k}$ and obtain the following lower bound
\begin{subequations}\label{eq:dmin_K1}
\begin{align}
&d_{\min}\left(\sum^K_{i=\ell_k}h_{i}\msf{X}_{i,\ell_k}\right) \nonumber \\
\overset{\eqref{eq:Xk_QAM}}{=}& d_{\min}\left(\sum^K_{i=\ell_k}h_{i}\eta_{i}\sqrt{P_i}2^{\frac{n_{\ell_k}-\log\SNR_i+\sum^K_{i'=i+1}m_{i',\ell_k}}{2}}\msf{F}_{i,\ell_k}\right) \\
\geq&
d_{\min}\left(\sum^K_{i=\ell_k} \sqrt{\frac{3}{2^{n_{\ell_k}}}}2^{\frac{n_{\ell_k}+\sum^K_{i'=i+1}m_{i',\ell_k}}{2}}\msf{F}_{i,\ell_k}\right) \label{eq:dmin_etaK} \\
=&\sqrt{3}\cdot  d_{\min}\left(\sum^K_{i=\ell_k} 2^{\frac{\sum^K_{i'=i+1}m_{i',\ell_k}}{2}}\msf{F}_{i,\ell_k}\right)\\
= &\sqrt{3}, \label{eq:dmin_K}
\end{align}
\end{subequations}
where in \eqref{eq:dmin_etaK} we have used \eqref{eq:eta_nk2} together with the fact that
\begin{subequations}
\begin{align}
E_{k,\ell_k}  \triangleq & \E\left[\left|2^{\frac{n_{\ell_k}-\log\SNR_k+\sum^K_{i=k+1}m_{i,\ell_k}}{2}}\msf{F}_{k,\ell_k}\right|^2\right] \\
=& \frac{2^{n_{\ell_k}-\log\SNR_k+\sum^K_{i=k+1}m_{i,\ell_k}}(2^{m_{k,\ell_k}}-1)}{6} \\
\leq&  \frac{2^{n_{\ell_k}-\log\SNR_k+\sum^K_{i=k}m_{i,\ell_k}}}{6} \\
\leq & \frac{2^{n_{\ell_k}-n_k+1+\sum^K_{i=k}m_{i,\ell_k}}}{6} \label{eq:eta_nk} \\
\leq & \frac{2^{n_{\ell_k}}}{3}, \label{eq:eta_nk1}
\end{align}
\end{subequations}
with \eqref{eq:eta_nk} due to $n_k \leq \log \SNR_k+1$ and recall that $\SNR_k=P_k|h_k|^2$, and \eqref{eq:eta_nk1} following from the inequality in \eqref{eq:rate_region_dK}. Moreover, \eqref{eq:dmin_K} follows that $d_{\min}(\sum^K_{i=\ell_k} 2^{\frac{\sum^K_{i'=i+1}m_{i',\ell_k}}{2}}\msf{F}_{i,\ell_k})=1$ by applying Lemma \ref{lem:dmin2} in Appendix \ref{sec:app}. From \eqref{eq:dmin_K1}, it can be seen that the proposed constellation and power coefficient designs in \eqref{eq:Xk_QAM} guarantees that the superimposed constellation $\sum^K_{i=\ell_k}h_{i}\msf{X}_{i,\ell_k}$ has a constant minimum distance lower bound. This is crucial for the proposed scheme to achieve close-to-capacity performance with TIN decoding as shown next.

Note also that according to Lemma \ref{lem:dmin2} in Appendix \ref{sec:app}, $\sum^K_{i=\ell_k} 2^{\frac{\sum^K_{i'=i+1}m_{i',\ell_k}}{2}}\msf{F}_{i,\ell_k}$ forms a regular QAM with zero mean and cardinality $2^{\sum^K_{i=\ell_k}m_{i,\ell_k}}$. In this case, we can lower bound the first term in \eqref{eq:Ik_first_term} as
\begin{subequations}\label{eq:Isupk}
\begin{align}
\mathds{I}&\left(\sum^K_{i=\ell_k}h_{i}\msf{X}_{i,\ell_k};\msf{Y}_{\ell_k}\right)\geq \mathbb{H}\left(\sum^K_{i=\ell_k}h_{i}\msf{X}_{i,\ell_k}\right)-\log\left(\frac{2\pi e}{12}\right) \nonumber\\
&-\log\left(1+\frac{12}{d^2_{\min}\left(\sum^K_{i=\ell_k}h_{i}\msf{X}_{i,\ell_k}\right)}\right),\label{eq:Isupk_1}\\
&\overset{\eqref{eq:dmin_K1}}{\geq} \sum^K_{i=\ell_k}m_{i,\ell_k}-\log\left(\frac{5\pi e}{6}\right),
\end{align}
\end{subequations}
where \eqref{eq:Isupk_1} follows the same line of approaches in \cite[Prop. 1]{7451210} by extending the arguments for a one-dimensional (not necessarily regular) PAM constellation therein to the two-dimensional QAM constellation in our case.

Substituting \eqref{eq:Isupk} into \eqref{eq:Ik_mutual_inf}, we get
\begin{align}\label{eq:IK_lower}
\mathds{I}(\msf{X}_{k,\ell_k};\msf{Y}_{\ell_k}) \geq m_{k,\ell_k}-\log\left(\frac{5\pi e}{6}\right),\;\ell_k\in\{1,\ldots,k\}.
\end{align}
Recall that the deterministic rate region $(m_{\ell_k,\ell_k},\ldots,m_{K,\ell_k})$ is within 1 bit to the capacity region of the corresponding Gaussian MAC \cite{Avestimehr11}. Thus, the proposed scheme with TIN decoding is capable of achieving the capacity region of the Gaussian MAC to within a \emph{constant gap independent of} the number of users and channel parameters according to \eqref{eq:IK_lower}. In fact, one can see from \eqref{eq:Isupk} that larger $d_{\min}(\sum^K_{i=\ell_k}h_{i}\msf{X}_{i,\ell_k})$ gives higher mutual information. However, instead of optimizing the minimum distance for a particular channel setting, the proposed scheme guarantees a constant minimum distance lower bound for universally achieving near-optimal performance.

\begin{remark}\label{remark:Gaussian}
One may want to know what would happen if all users adopt capacity-achieving Gaussian signaling while the receiver still performs TIN decoding. By analyzing the mutual information from \eqref{eq:Ik_mutual_inf_step1}-\eqref{eq:Ik_mutual_inf_step3}, for user $k\in\{1,\ldots,K\}$, we have
\begin{align}
\mathds{I}(\msf{X}_{k,\ell_k};\msf{Y}_{\ell_k}) =& \log\left(1+\frac{\SNR_k}{1+\sum_{i\in\{\ell_k,\ldots,K\}\setminus k}\SNR_i}\right),\nonumber \\
\ell_k \in&\{1,\ldots,k\}.
\end{align}
Under the channel ordering $\SNR_1>\ldots>\SNR_K$, we have $\mathds{I}(\msf{X}_{k,\ell_k};\msf{Y}_{\ell_k})<1$ for $\ell_k \in\{1,\ldots,k-1\}$ because $\ell_k<k$ leads to $\SNR_k<\SNR_{\ell_k}$. Clearly, Gaussian signaling with TIN decoding performs poorly and is strictly suboptimal in this case. This is because Gaussian interference lacks structures which may be difficult to exploit in TIN decoding.
\demo
\end{remark}

\begin{remark}\label{remark:power_ratio}
An interesting feature of the proposed scheme is that the close-to-capacity mutual information in \eqref{eq:IK_lower} is achieved without the need of using full transmit power. To see this, we compute the ratio between the actual average power and the individual power constraint for user $k$ as
\begin{subequations}\label{eq:power_ratio}
\begin{align}
 \zeta_{k,\ell_k}\triangleq &\frac{\E[|\msf{X}_{k,\ell_k}|^2]}{P_k}= \frac{E_{k,\ell_k}}{\max\{E_{\ell_k,\ell_k},\ldots,E_{K,\ell_k}\}}\\
 =& \frac{2^{n_{\ell_k}-\log \SNR_k+\sum^K_{i=k+1}m_{i,\ell_k}}(2^{m_{k,\ell_k}}-1)}{2^{n_{\ell_k}-\log \SNR_{k'}+\sum^K_{i=k'+1}m_{i,\ell_k}}(2^{m_{k',\ell_k}}-1)} \label{eq:power_ratio1} \\
\geq& \frac{2^{-n_k+\sum^K_{i=k+1}m_{i,\ell_k}}(2^{m_{k,\ell_k}}-1)}{2^{-n_{k'}+1+\sum^K_{i=k'+1}m_{i,\ell_k}}(2^{m_{k',\ell_k}}-1)}  \label{eq:power_ratio2} \\
=&\frac{2^{-m_{k,\ell_k}}(2^{m_{k,\ell_k}}-1)}{2^{1-m_{k',\ell_k}}(2^{m_{k',\ell_k}}-1)} \label{eq:power_ratio3}\\
\geq& \frac{1}{4}, \label{eq:power_ratio4}
\end{align}
\end{subequations}
where in \eqref{eq:power_ratio1} we have let $E_{k',\ell_k}=\max\{E_{\ell_k,\ell_k},\ldots,E_{K,\ell_k}\}$ for some $k'\in\{\ell_k,\ldots,K\}$, \eqref{eq:power_ratio2} follows from the facts that $n_{k'}-1<\log\SNR_{k'}$ and $\log\SNR_{k}<n_k$, \eqref{eq:power_ratio3} follows by using the equality of \eqref{eq:rate_region_dK}, and \eqref{eq:power_ratio4} follows that $m_{k,\ell_k}\geq 1$ and $m_{k',\ell_k}<\infty$. Hence, the average transmit power saving for user $k$ can be up to $75\%$. When only the strict inequality of \eqref{eq:rate_region_dK} holds, the average power saving can be larger than 75\% as will be shown in the numerical results. However, the exact power-saving ratio for this case is difficult to derive.
\demo
\end{remark}

Having derived the mutual information of the proposed scheme with TIN decoding, we will then analyze the finite blocklength achievable rate, which is dominant by the first-order term, i.e., mutual information, and second-order term, dispersion, as well as decoding error probability and blocklength.

\subsection{Finite Blocklength Coding Analysis and Design}\label{sec:FBL}

In this section, we present the finite blocklength achievable rate of the $K$-user heterogeneous MAC with an arbitrary choice of signaling constellations, lengths of transmitted blocks, and error probability requirements.

We adopt the same notations as introduced in the second paragraph of Section \ref{sec:analysis}. The main result of this section is stated in Theorem \ref{prof:u1}.

\begin{theorem}\label{prof:u1}
Let $\epsilon_k$ be the upper bound on the average TIN decoding error probability of user $k$. For the $K$-user MAC with heterogeneous blocklength constraints $(N_1,\ldots,N_k)$ and error probability constraints $(\epsilon_1,\ldots,\epsilon_K)$, the rate of user $k$ with TIN decoding, i.e., $R_k$, is achievable if it satisfies
\begin{align}
R_k \leq& \sum^k_{\ell_k=1}\frac{N_{\ell_k}-N_{\ell_k-1}}{N_k}\mathds{I}(X_{k,\ell_k};Y_{\ell_k})\nonumber \\
&-\frac{\sqrt{\sum^k_{\ell_k=1}(N_{\ell_k}-N_{\ell_k-1})\mathbb{V}(X_{k,\ell_k};Y_{\ell_k})}}{N_k}Q^{-1}\left(\epsilon_k\right) \nonumber\\
&+O\left(\frac{1}{N_k}\right), \label{eq:u1rate}
\end{align}
where
\begin{align}
\mathds{I}(\msf{X}_{k,\ell_k};\msf{Y}_{\ell_k}) =& \E[i(\msf{X}_{k,\ell_k};\msf{Y}_{\ell_k})], \\
\mathbb{V}(\msf{X}_{k,\ell_k};\msf{Y}_{\ell_k}) =& \text{Var}[i(\msf{X}_{k,\ell_k};\msf{Y}_{\ell_k})],
\end{align}
are the mutual information and dispersion, respectively, and $Q^{-1}(x)$ is the inverse of $Q$ function $Q(x)=\int^{\infty}_{x}\frac{1}{\sqrt{2\pi}}e^{-\frac{t^2}{2}}dt$. The information density $i(\msf{X}_{k,\ell_k};\msf{Y}_{\ell_k})$ is given in \eqref{eq:infor_den} on the top of page 10.
\end{theorem}

\begin{figure*}[t]
\begin{align}\label{eq:infor_den}
i(\msf{X}_{k,\ell_k};\msf{Y}_{\ell_k}) =  \log \left( \frac{\sum\limits_{x_{\ell_k,\ell_k},\ldots,x_{k-1,\ell_k},x_{k+1,\ell_k},\ldots,x_{K,\ell_k}}e^{-\left|y_{\ell_k}-\sum^K_{k'=\ell_k}h_{k'}x_{k',\ell_k}\right|}}{\frac{1}{|\Lambda_{k,\ell_k}|}\sum\limits_{x_{\ell_k,\ell_k},\ldots,x_{K,\ell_k}}e^{-\left|y_{\ell_k}-\sum^K_{k'=\ell_k}h_{k'}x_{k',\ell_k}\right|}}\right).
\end{align}
\end{figure*}

\begin{IEEEproof}
The proof is given in Appendix \ref{sec:app2}.
\end{IEEEproof}

\begin{remark}
The impacts of interfering symbol lengths on the achievable rate are clearly shown in \eqref{eq:u1rate}. Under heterogeneous interference, the achievable rate is no longer determined by a single SINR as in the homogeneous blocklength case. In addition, the relationship between heterogeneous blocklengths and achievable rate in \eqref{eq:u1rate} may also hold for other i.i.d. inputs, e.g., i.i.d. Gaussian codes. However, for the shell codes commonly used for deriving larger finite blocklength achievable rates in many works \cite{5452208,7300429,9535162}, this relation may no longer hold as the distribution of each symbol of shell codes is not independent of each other.
\demo
\end{remark}

From Theorem \ref{prof:u1}, we note that for given power, blocklength, and error probability constraints, larger mutual information and smaller dispersion are desirable for achieving a larger rate.

With the derived achievable rate, we can determine the modulation and channel coding parameters for the proposed schemes in the following steps.
\begin{enumerate}
\item Find the modulation orders $(m_{\ell_k,\ell_k},\ldots,m_{K,\ell_k})$ that satisfy \eqref{eq:rate_region_dK}, for $\ell_k\in\{1,\ldots,k\}$ and $k\in\{1,\ldots,K\}$.

\item For user $k$, given symbol length $N_k$ and error probability $\epsilon_k$ requirements, compute the achievable rate $R_k$ using \eqref{eq:u1rate} for the proposed signaling $\msf{X}_k$ in \eqref{eq:Xk_QAM}.

\item For user $k$, match the achievable rate with the transmission rate such that $R_k = \frac{I_k}{L_k}(\sum^k_{\ell_k=1}\frac{N_{\ell_k}-N_{\ell_k-1}}{N_k}m_{k,\ell_k})$. By \eqref{eq:Lk_Nk}, we obtain the information and codeword lengths as $(I_k,L_k) = (R_kN_k,\sum^k_{\ell_k=1}(N_{\ell_k}-N_{\ell_k-1})m_{k,\ell_k})$.
\end{enumerate}
Once the parameters have been determined, user $k$ can then pick a point-to-point binary channel code $\mathcal{C}_k(I_k,L_k)$ and use QAM modulation orders $(m_{k,1},\ldots,m_{k,k})$ to achieve its target rate $R_k$ with given blocklength $N_k$ and error probability $\epsilon_k$ constraints.

\section{Numerical Results}
We provide numerical results in terms of finite blocklength achievable rates, dispersion, and error probability of a practical set-up by adopting off-the-shelf channel codes.

\subsection{Numerical Results on Achievable Rates}\label{sec:sim1}
\begin{figure}[t!]
	\centering
\includegraphics[width=\linewidth]{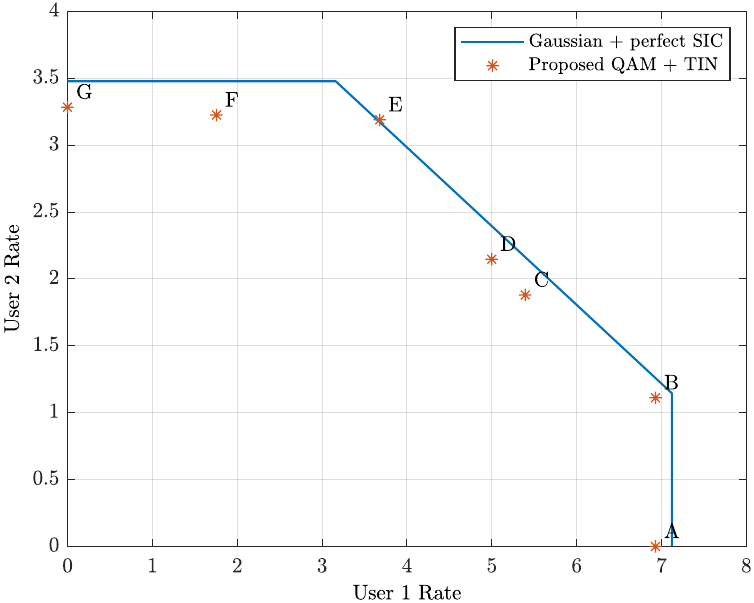}
\caption{Achievable rate pairs of two users in bits/s/Hz.}
\label{fig:rate1}
\end{figure}

\begin{figure}[t!]
	\centering
\includegraphics[width=\linewidth]{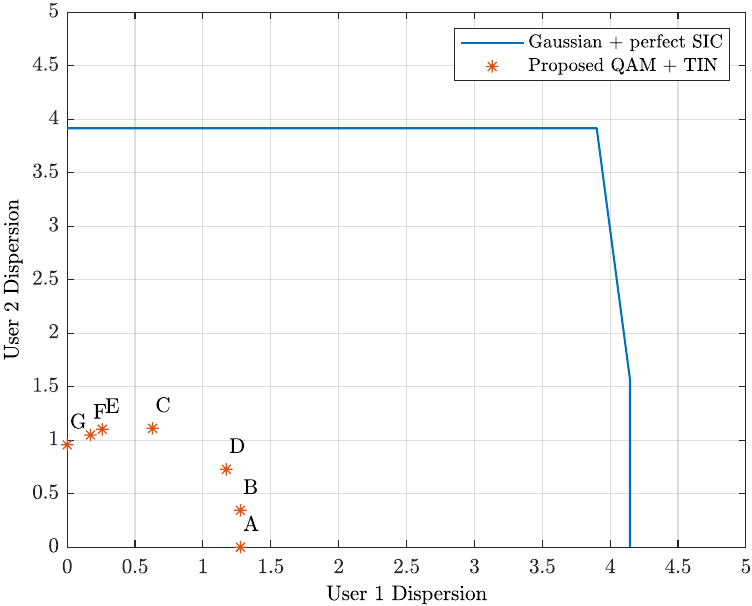}
\caption{Dispersion pairs of two users in bits$^2$/s/Hz.}
\label{fig:dispersion1}
\end{figure}

In this section, we provide numerical results to show the finite blocklength achievable rates of the proposed schemes using QAM with TIN. Specifically, two schemes are evaluated, where the coded modulation scheme translated from the deterministic scheme based on \eqref{Eq:GK} is referred to as the type I scheme and the one based on \eqref{Eq:GKa} and \eqref{Eq:GKb} is the type II scheme. We consider $(\SNR_1,\SNR_2)=(24,12)$ dB, $(\epsilon_1,\epsilon_2)=(10^{-6},10^{-5})$, and $(N_1,N_2)=(128,200)$. The achievable rate pairs and the corresponding dispersion pairs are shown in Figs. \ref{fig:rate1} and \ref{fig:dispersion1}, respectively, while their scheme types, modulation orders $m$ and power consumption ratios $\zeta$ (see Remark \ref{remark:power_ratio}) are reported in Table \ref{tab:num1}. For illustration purposes, we follow the 5G modulation standard in \cite{TS138212_v16p8}, such that the modulation orders of the individual and superimposed constellations are even numbers.
\begin{table}[h]
    \centering
       \begin{threeparttable}
     \caption{Scheme types, modulation orders, and power savings for the rate points in Fig. \ref{fig:rate1} and their corresponding dispersion points in Fig \ref{fig:dispersion1}.}
    \label{tab:num1}
    \begin{tabular}{|c|c|c|c|c|c|c|c|}
        \hline Point & A & B & C & D & E &F  &G \\
        \hline Scheme & I\&II &I\&II & I & II & I\&II & I\&II & I\&II \\
        \hline $m_1$     & 8 & 8 & 6 & 6 & 4 & 2 & 0 \\
        \hline $m_{2,1}$ & 0 & 0 & 2 & 2 & 4 & 4 & 4 \\
        \hline $m_{2,2}$ & 0 & 4 & 4 & 4 & 4 & 4 & 4 \\
        \hline $\zeta_1$ & 1 & 1 & 1 & 1 & 1 & 0.202 & 0 \\
        \hline $\zeta_{2,1}$ & 0 & 1 & 0.189 & 0.783 & 0.991 & 1 & 1 \\
        \hline
    \end{tabular}
    \begin{tablenotes}
      \scriptsize
      \item Note: I\&II means that both schemes achieve the same rate. $\zeta_{2,2}=1$ for all points.
    \end{tablenotes}
      \end{threeparttable}
\end{table}

For comparison purposes, we include the performance of a benchmark scheme based on Gaussian signaling and perfect SIC. By perfect SIC we mean that the decoding and cancellation of another user's interfering signal is always feasible and successful such that the decoding error probability is only associated with the decoding of a user's own signal. Since the characterization of the rate region of such a benchmark scheme is missing for the considered channel, we optimistically assume that the rate region is generated by using the convex combination of all corner points as in the conventional GMAC. It should be noted, however, that under finite blocklength, SIC can fail and cause error propagation while time-sharing is not able to provide a convex region \cite{6665138}. Moreover, it should also be kept in mind that in the heterogeneous case, the messages of the more urgent user have to be decoded without waiting for the reception of the whole transmission block of the other user; therefore, SIC is infeasible in URLLC communication scenarios. Thus, the rate region of the benchmark scheme is a hypothetical rate region used for comparison purposes only, which can be regarded as an outer bound of the actual rate region under imperfect SIC with a non-convex time-sharing region.

Fig. \ref{fig:rate1} shows that the proposed scheme with QAM and TIN can perform very close to the benchmark scheme based on Gaussian signaling and perfect SIC. Intuitively, this is due to exploiting the structural interference introduced by our carefully designed discrete input signaling. It is also important to emphasize that such performance is achieved without the need to use full transmit power as shown in Table \ref{tab:num1}. Further, we note that point E is slightly outside the rate region of the benchmark scheme, implying that the proposed schemes can allow both users to achieve rates larger than Gaussian signaling with perfect SIC decoding. Fundamentally, this is due to close-to-capacity mutual information and much smaller dispersion of the proposed scheme compared to the benchmark scheme, as revealed in Section \ref{sec:analysis} and Fig. \ref{fig:dispersion1}, respectively. We highlight that the main features of URLLC communications are short blocklength and ultra-low error probability. In this scenario, a small increase in the dispersion can have a non-negligible impact on the finite blocklength achievable rate. On the other hand, by decreasing the target error probability while keeping other parameters fixed, the achievable rate of the proposed scheme could eventually surpass that of the benchmark scheme. This is because the second term of the proposed scheme in \eqref{eq:u1rate} grows slower than that of Gaussian signaling due to smaller dispersion as the error probability decreases.

\subsection{Numerical Results on Error Probability}
\begin{figure}[t!]
	\centering
\includegraphics[width=\linewidth]{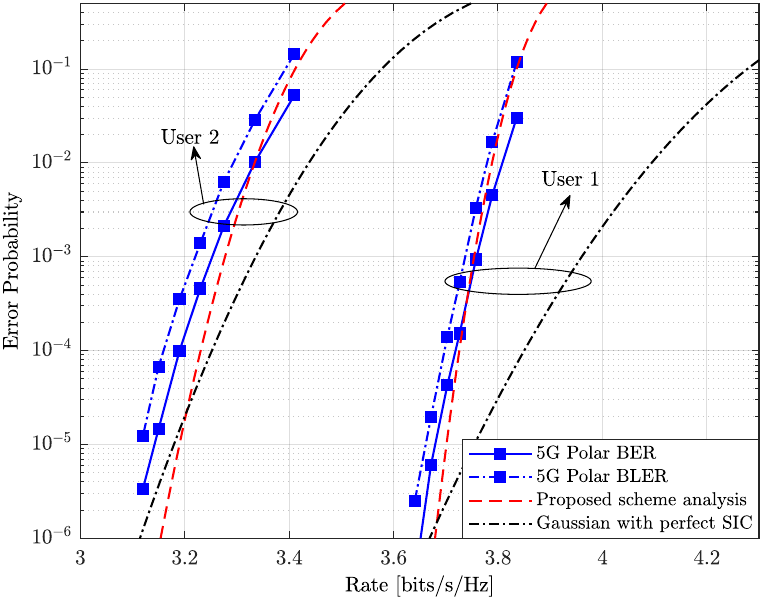}
\caption{Error probability versus the finite blocklength achievable rate for both users.}
\label{fig:u1err}
\end{figure}

We evaluate the error performance of the proposed scheme by employing off-the-shelf channel codes. As in Section \ref{sec:sim1}, we consider $(\SNR_1,\SNR_2)=(24,12)$ dB, $(N_1,N_2)=(128,200)$, and pick $(m_1,m_{2,1},m_{2,2})=(4,4,4)$ according to Table \ref{tab:num1}. Then, given a range of target error probabilities $(\epsilon_1,\epsilon_2)$, we obtain the corresponding achievable rates $(R_1,R_2)$ by \eqref{eq:u1rate}. As a result, the information and codeword lengths of the channel codes for both users are determined by using the methods in the last two paragraphs of Section \ref{sec:FBL}. Both users employ the 5G standard polar codes with 11-bit cyclic redundancy check (CRC) from \cite{TS138212_v16p8} and use the puncturing and shortening steps therein. For bit mapping, we consider random interleavers, i.e., random for each channel realization. Moreover, the decoding of users 1 and 2's messages is accomplished by adaptive successive-cancellation list decoding \cite{6355936} with the maximum list sizes 128 and 1024, respectively. The bit error rate (BER) and block error rate (BLER) of both users versus the rate are reported in Fig. \ref{fig:u1err}. For comparison purposes, we include $(\epsilon_1,\epsilon_2)$ of the benchmark scheme based on Gaussian signaling and perfect SIC decoding.

Observe that the BER and BLER of both users closely match with the analytical average block error probability upper bounds of the proposed scheme. This confirms the accuracy of the derived finite blocklength achievable rate. In addition, each user's achievable rate is very close to that of Gaussian signaling with perfect SIC decoding when the required error probability is low. We emphasize that such performance is achieved using existing off-the-shelf binary codes without explicitly optimizing the underlying code structures. This further demonstrates the effectiveness of the proposed scheme when using practical coding, modulations, and low-complexity single-user decoding. We expect that the proposed scheme may surpass the Gaussian benchmark scheme at low error probability with a tailored coding design and iterative detection and decoding.

In light of the above numerical results in terms of achievable rates and error probabilities, we believe that the proposed scheme is promising for supporting simultaneous communications of heterogeneous (URLLC) services in the uplink.

\section{Conclusion}
In this paper, we investigated the uplink multiple access of heterogeneous services with different blocklength and error probability constraints. We first approximated the channel model as a deterministic channel and constructed capacity-achieving coding schemes with TIN decoding. Then, we translated the aforementioned coding schemes into the coded modulation schemes with TIN decoding for the homogeneous MAC. The proposed scheme enjoys several desired properties, such as single-user encoding and decoding complexities as well as average transmit power saving. We analyzed the mutual information achieved by the proposed scheme and proved that its gap to capacity is upper bounded by a constant independent of the number of users and channel coefficients. Moreover, we derived the finite blocklength achievable rate for the $K$-user heterogeneous MAC with discrete input distributions and TIN decoding. Numerical results demonstrated that the proposed scheme using QAM and TIN decoding can operate close to or even outperform the benchmark scheme based on capacity-achieving Gaussian signaling and perfect SIC decoding (infeasible in practice).

\appendices
\section{Useful Lemmas for Superimposed Constellations}\label{sec:app}
\begin{lemma}\label{lem:dmin}
Consider a pair of regular QAM constellations $(\Lambda_1,\Lambda_2)$ with zero mean and minimum distances satisfy $d_{\min}(\Lambda_1) = d_{\min}(\Lambda_2)=\delta>0$. The superimposed constellation $\Lambda_1+2^{\frac{\log|\Lambda_1|}{2}}\Lambda_2$ is a regular QAM with zero mean, $d_{\min}(\Lambda_1+2^{\frac{\log|\Lambda_1|}{2}}\Lambda_2) = \delta$, and cardinality $|\Lambda_1|\cdot|\Lambda_2|$.
\end{lemma}
\begin{IEEEproof}
We first look at the superposition of $\Lambda_1$ and $\Lambda_2$ in real parts by fixing the imaginary parts, i.e., $\Re(\Lambda_1)+2^{\frac{\log|\Lambda_1|}{2}}\Re(\Lambda_2)$. For $k\in\{1,2\}$, we have
\begin{align}\label{eq:PAM_set}
\Re(\Lambda_k)=\left\{-\frac{2^{\frac{\log|\Lambda_k|}{2}-1}}{2}\delta, -\frac{2^{\frac{\log|\Lambda_k|}{2}-3}}{2}\delta,\ldots,\frac{2^{\frac{\log|\Lambda_k|}{2}-1}}{2}\delta\right\}.
\end{align}
Clearly, for any two neighboring constellation points $\lambda_{k,1},\lambda_{k,2} \in \Re(\Lambda_k)$ and $\lambda_{k,1}<\lambda_{k,2}$, we have
\begin{align}\label{eq:L2_dmin}
\lambda_{k,2}-\lambda_{k,1} = \delta.
\end{align}
Next, we treat $\Re(\Lambda_1)$ as a cluster and compute the inter-cluster distance of $\Re(\Lambda_1)+2^{\frac{\log|\Lambda_1|}{2}}\Re(\Lambda_2)$ as
\begin{subequations}\label{eq:dcl_step2}
\begin{align}
d(\Re&(\Lambda_1)+\lambda_{2,1},\Re(\Lambda_1)+\lambda_{2,2})=-\min\{\Re(\Lambda_1)\}\nonumber \\
=&+2^{\frac{\log|\Lambda_1|}{2}}\lambda_{2,2}
-\max\{\Re(\Lambda_1)\}-2^{\frac{\log|\Lambda_1|}{2}}\lambda_{2,1} \\
=&-\frac{2^{\frac{\log|\Lambda_1|}{2}-1}}{2}\delta\cdot2+2^{\frac{\log|\Lambda_1|}{2}}\delta \label{eq:dcl_step1} \\
=& \delta,
\end{align}
\end{subequations}
where \eqref{eq:dcl_step1} follows by using \eqref{eq:PAM_set} with $k=1$ and \eqref{eq:L2_dmin} with $k=2$.
According to \eqref{eq:L2_dmin} and \eqref{eq:dcl_step2}, we know that for any pair of neighboring points $\lambda'_1,\lambda'_2\in\Re(\Lambda_1)+2^{\frac{\log|\Lambda_1|}{2}}\Re(\Lambda_2)$ and $\lambda'_1<\lambda'_2$, we have
\begin{align}
\lambda'_2-\lambda'_1=\delta.
\end{align}
As a result, the following holds.
\begin{align}\label{eq:sup_real}
\Re(\Lambda_1)&+2^{\frac{\log|\Lambda_1|}{2}}\Re(\Lambda_2) = \left\{-\frac{2^{\frac{\log(|\Lambda_1|\cdot|\Lambda_2|)}{2}-1}}{2}\delta, \right. \nonumber \\
&\left.-\frac{2^{\frac{\log(|\Lambda_1|\cdot|\Lambda_2|)}{2}-3}}{2}\delta,\ldots, \frac{2^{\frac{\log(|\Lambda_1|\cdot|\Lambda_2|)}{2}-1}}{2}\delta\right\}.
\end{align}
Due to the symmetry property of the regular QAM, we also have that
\begin{align}\label{eq:sup_imag}
\Im(\Lambda_1)+2^{\frac{\log|\Lambda_1|}{2}}\Im(\Lambda_2) =\Re(\Lambda_1)&+2^{\frac{\log|\Lambda_1|}{2}}\Re(\Lambda_2).
\end{align}
Combining \eqref{eq:sup_real} and \eqref{eq:sup_imag}, we arrive at the conclusion stated in Lemma \ref{lem:dmin}.
\end{IEEEproof}

\begin{lemma}\label{lem:dmin2}
Consider $K$ sets of regular QAM constellations $(\Lambda_1,\ldots,\Lambda_K)$ with zero mean and minimum distances satisfy $d_{\min}(\Lambda_k) =\delta>0,\forall k\in\{1,\ldots,K\}$. The superimposed constellation $\Lambda = \Lambda_1+\sum^K_{k=2}2^{\frac{\sum^{k-1}_{i=1}\log|\Lambda_i|}{2}}\Lambda_k$ is a regular QAM with zero mean, $d_{\min}(\Lambda) = \delta$, and cardinality $|\Lambda| = \prod^K_{k=1}|\Lambda_k|$.
\end{lemma}
\begin{IEEEproof}
The proof follows by recursively applying Lemma \ref{lem:dmin} on $\Lambda$ for $K-1$ times.
\end{IEEEproof}

\section{Proof of Theorem \ref{prof:u1}}\label{sec:app2}

By Theorem 17 of \cite{5452208}, there exists an a length-$N_k$ code $\mathcal{C}_k$ whose average decoding error probability as a function of $N_k$ is upper bounded by
\begin{align}\label{eq:RCU}
\epsilon_k(N_k) \leq \E\left[2^{-\max\left\{0,i(\msf{X}^{[N_k]};\msf{Y}^{[N_k]})-\log\frac{M_k-1}{2}\right\}}\right],
\end{align}
where $M_k$ is the size of codebook $\mathcal{C}_k$. Next, we need to show that the following holds for some $\alpha>0$
\begin{subequations}\label{eq:dt0}
\begin{align}
\epsilon_k(N_k)& \geq\E\left[2^{-\max\{0,i(\msf{X}^{[N_k]};\msf{Y}^{[N_k]})-\log \alpha\} }\right]\label{eq:dt2} \\
=& \mathbb{P}\left[i\left(\msf{X}^{[N_k]};\msf{Y}^{[N_k]}\right)\leq \log \alpha\right] \nonumber \\
&+\alpha \E\left[2^{i(\msf{X}^{[N_k]};\msf{Y}^{[N_k]})}\mathds{1}_{\{i(\msf{X}^{[N_k]};\msf{X}^{[N_k]})>\log \alpha\}} \right], \label{eq:dt1}
\end{align}
\end{subequations}
where $\mathds{1}_{\{ .\}}$ denotes the indicator function. Note that $\msf{X}_k[i]$ and $\msf{X}_k[j]$ are independent and $\msf{Y}_k[i]$ and $\msf{Y}_k[j]$ are independent for any $i\neq j$ and $i,j\in\{1,\ldots,N_k\}$. By the definition of information density \cite{5452208}, we have
\begin{align}
i\left(\msf{X}_k^{[N_k]};\msf{Y}_k^{[N_k]}\right)=\sum^{N_k}_{j=1}i(\msf{X}_k[j];\msf{Y}_k[j]).
\end{align}
As a result, the mutual information, i.e., the mean of information density, satisfies
\begin{subequations}\label{eq:Ik_decompose}
\begin{align}
\mathds{I}\left(\msf{X}^{[N_k]}_{k};\msf{Y}^{[N_k]}_{k}\right) &= \E\left[i\left(\msf{X}^{[N_k]}_{k};\msf{Y}^{[N_k]}_{k}\right)\right]\\
=&\sum^{N_k}_{j=1}\E[i(\msf{X}_{k}[j];\msf{Y}_{k}[j])] \\
=&\sum^k_{\ell_k=1}(N_{\ell_k}-N_{\ell_k-1})\E[i(X_{k,\ell_k};Y_{\ell_k})]\label{eq:Ink_xkk}\\
=&\sum^k_{\ell_k=1}(N_{\ell_k}-N_{\ell_k-1})\mathds{I}(X_{k,\ell_k};Y_{\ell_k}).
\end{align}
\end{subequations}
Similarly, the variance of information density satisfies
\begin{subequations}\label{eq:Vk_decompose}
\begin{align}
\mathbb{V}\left(\msf{X}^{[N_k]}_{k};\msf{Y}^{[N_k]}_{k}\right) &= \sum^{N_k}_{j=1}\text{Var}[i(\msf{X}_{k}[j];\msf{Y}_{k}[j])] \\
=\sum^k_{\ell_k=1}&(N_{\ell_k}-N_{\ell_k-1})\text{Var}[(\msf{X}_{k,\ell_k};\msf{Y}_{\ell_k})]\\
=\sum^k_{\ell_k=1}&(N_{\ell_k}-N_{\ell_k-1})\mathbb{V}(\msf{X}_{k,\ell_k};\msf{Y}_{\ell_k}).
\end{align}
\end{subequations}

We can first bound the second term in \eqref{eq:dt1} by directly using Lemma 47 from \cite{5452208} without the need of determining the value of $\alpha$
\begin{align}
\alpha& \E\left[2^{i(\msf{X}^{[N_k]};\msf{Y}^{[N_k]})}\mathds{1}_{\{i(\msf{X}^{[N_k]};\msf{X}^{[N_k]})>\log \alpha\}} \right] \nonumber \\
&\leq \frac{2}{\sqrt{2\pi\sum^k_{\ell_k=1}(N_{\ell_k}-N_{\ell_k-1})\mathbb{V}(\msf{X}_{k,\ell_k};\msf{Y}_{\ell_k})}}+\frac{4B_k}{\sqrt{N_k}}. \label{eq:second_term}
\end{align}
where $B_k = \frac{C_0\sum^k_{\ell_k=1}\frac{N_{\ell_k}-N_{\ell_k-1}}{N_k}\mathbb{E}[|i(\msf{X}_{k,\ell_k};\msf{Y}_{\ell_k})-I(\msf{X}_{k,\ell_k};\msf{Y}_{\ell_k})|^3]}{\left(\sum^k_{\ell_k=1}\frac{N_{\ell_k}-N_{\ell_k-1}}{N_k}\mathbb{V}(\msf{X}_{k,\ell_k};\msf{Y}_{\ell_k})\right)^\frac{3}{2}}$ is a constant since $C_0=0.5600$ \cite{Shevtsova2010} and $\msf{X}_{k,\ell_k}$ is uniformly distributed over a finite constellation set $\Lambda_{k,\ell_k}$.

To bound the first term in \eqref{eq:dt1}, we let
\begin{subequations}\label{eq:log_alpha}
\begin{align}
\log \alpha =& \mathds{I}\left(\msf{X}^{[N_k]}_{k};\msf{Y}^{[N_k]}_{k}\right)-\lambda_k\sqrt{\mathbb{V}\left(\msf{X}^{[N_k]}_{k};\msf{Y}^{[N_k]}_{k}\right)} \\
=& \sum^k_{\ell_k=1}(N_{\ell_k}-N_{\ell_k-1})\mathds{I}(X_{k,\ell_k};Y_{\ell_k}) \nonumber \\
&-\lambda_k \sqrt{\sum^k_{\ell_k=1}(N_{\ell_k}-N_{\ell_k-1})\mathbb{V}(\msf{X}_{k,\ell_k};\msf{Y}_{\ell_k})} ,
\end{align}
\end{subequations}
for arbitrary $\lambda_k$. By doing so, we can use the Berry-Esseen central limit Theorem \cite[Th. 2, Ch. XVI-5]{Feller_book} to bound the probability in the first term of \eqref{eq:dt1} as
\begin{align}
&\left|\mathbb{P}\left[\frac{\sum^{N_k}_{j=1}\left(i(\msf{X}[j];\msf{Y}[j])-\mathds{I}(\msf{X}[j];\msf{Y}[j])\right)}{\sqrt{\sum^{N_k}_{j=1}\mathbb{V}(\msf{X}[j];\msf{Y}[j])}}\leq \lambda_k\right]-Q(\lambda_k)\right|\nonumber \\
 &\leq \frac{B_k}{\sqrt{N_k}}, \\
\Rightarrow & \mathbb{P}\left[i\left(\msf{X}^{[N_k]};\msf{Y}^{[N_k]}\right)\leq \log \alpha\right] \leq Q(\lambda_k)+\frac{B_k}{\sqrt{N_k}},\label{eq:BMCT}
\end{align}
where \eqref{eq:BMCT} follows by using \eqref{eq:Ik_decompose}, \eqref{eq:Vk_decompose}, and \eqref{eq:log_alpha}. To determine $\lambda_k$, we make the sum of \eqref{eq:second_term} and \eqref{eq:BMCT} equal to $\epsilon_k(N_k)$ in order to fulfill \eqref{eq:dt1}. Hence,
\begin{align}
\epsilon_k(N_k) =& \frac{2}{\sqrt{2\pi\sum^k_{\ell_k=1}(N_{\ell_k}-N_{\ell_k-1})\mathbb{V}(\msf{X}_{k,\ell_k};\msf{Y}_{\ell_k})}} \nonumber \\
&+Q(\lambda_k)+\frac{5B_k}{\sqrt{N_k}}.
\end{align}
This means that
\begin{align}\label{eq:lambda_k}
\lambda_k &= Q^{-1}\Bigg( \epsilon_k(N_k) \nonumber \\
&-\frac{2}{\sqrt{2\pi\sum^k_{\ell_k=1}(N_{\ell_k}-N_{\ell_k-1})\mathbb{V}(\msf{X}_{k,\ell_k};\msf{Y}_{\ell_k})}} -\frac{5B_k}{\sqrt{N_k}}\Bigg).
\end{align}
With $\lambda_k$ and $\alpha$ given in \eqref{eq:lambda_k} and \eqref{eq:log_alpha}, respectively, \eqref{eq:dt2} is proved by summing \eqref{eq:second_term} and \eqref{eq:BMCT}.

Combining both inequalities \eqref{eq:RCU} and \eqref{eq:dt0} for $\epsilon_k(N_k)$, we lower bound user $k$'s codebook size as
\begin{align}
&\log \frac{M_k-1}{2} \geq \log \alpha \\
\overset{\eqref{eq:log_alpha}}{\Rightarrow}& \log M_k \geq \sum^k_{\ell_k=1}(N_{\ell_k}-N_{\ell_k-1})\mathds{I}(X_{k,\ell_k};Y_{\ell_k}) \nonumber \\
&- \sqrt{\sum^k_{\ell_k=1}(N_{\ell_k}-N_{\ell_k-1})\mathbb{V}(\msf{X}_{k,\ell_k};\msf{Y}_{\ell_k})}Q^{-1}(\epsilon_k(N_k))\nonumber \\
&+O(1), \label{eq:taylor}
\end{align}
where in \eqref{eq:taylor} we have applied the first-order Taylor expansion of $Q^{-1}(.)$ as in \cite{5452208,JSACQiu22}. To ensure the TIN decoding error probability does not exceed $\epsilon_k$, i.e., $\epsilon_k(N_k) < \epsilon_k$, we rearrange \eqref{eq:taylor} into an upper bound of $\epsilon_k(N_k)$ and let its right-hand-side further upper bounded by $\epsilon_k$, such that
\begin{align}
&\epsilon_k(N_k) \nonumber \\
\leq& Q\left(\frac{\sum^k_{\ell_k=1}(N_{\ell_k}-N_{\ell_k-1})\mathds{I}(X_{k,\ell_k};Y_{\ell_k})-\log M_k +O(1)}{\sqrt{\sum^k_{\ell_k=1}(N_{\ell_k}-N_{\ell_k-1})\mathbb{V}(\msf{X}_{k,\ell_k};\msf{Y}_{\ell_k})}}  \right)\nonumber\\
\leq& \epsilon_k.
\end{align}
As a result, we obtain the upper bound of $R_k$ as
\begin{align}
R_k\leq & \frac{\log M_k}{N_k} \leq \sum^k_{\ell_k=1}\frac{N_{\ell_k}-N_{\ell_k-1}}{N_k}\mathds{I}(X_{k,\ell_k};Y_{\ell_k}) \nonumber \\
&- \frac{\sqrt{\sum^k_{\ell_k=1}(N_{\ell_k}-N_{\ell_k-1})\mathbb{V}(\msf{X}_{k,\ell_k};\msf{Y}_{\ell_k})}}{N_k}Q^{-1}(\epsilon_k)\nonumber \\
&+O\left(\frac{1}{N_k}\right). \label{eq:proof_rk}
\end{align}

Note that we obtain an improved achievability bound compared to the one in \cite{JSACQiu22}. This can be seen by noting that the last term of \eqref{eq:proof_rk} can be written as $-\frac{c_0}{N_k}$ for some positive constant $c_0$, whereas the last term of (44) in \cite{JSACQiu22} is $-\frac{c_1\log N_k}{N_k}$ for some positive constant $c_1$.

\bibliographystyle{myIEEEtran.bst}
\bibliography{MinQiu}

\end{document}